\DocumentMetadata{}
\documentclass[acmsmall,screen]{acmart}

\usepackage{url}
\usepackage{xspace}
\usepackage{wrapfig}
\usepackage{hyperxmp}
\usepackage{booktabs}
\usepackage{makecell}
\usepackage{hyperref}
\usepackage{graphicx}
\usepackage{colortbl}
\usepackage{multirow}
\usepackage{subfigure}
\usepackage{todonotes}
\usepackage{tablefootnote}
\usepackage{color, xcolor}
\usepackage{amsthm,amsmath,amsfonts}
\usepackage{enumitem}
\newcommand{\ie}{\textit{i.e.,}\xspace}
\newcommand{\eg}{\textit{e.g.,}\xspace}
\newcommand{\etc}{\textit{etc.}\xspace}
\newcommand{\etal}{\textit{et al.}\xspace}

\newcommand{\figref}[1]{Fig.~\ref{#1}\xspace}
\newcommand{\tabref}[1]{Table~\ref{#1}\xspace}
\newcommand{\secref}[1]{Section~\ref{#1}\xspace}

\newcommand{\papernum}{271\xspace}
\newcommand{\visnum}{92\xspace}
\newcommand{\gen}{GUI test generation\xspace} 
\newcommand{\rr}{GUI test record \& replay\xspace} 
\newcommand{\frm}{GUI testing framework\xspace} 
\newcommand{\srp}{GUI test script maintenance\xspace} 
\newcommand{\rpt}{GUI test report analysis\xspace} 
\newcommand{\ele}{GUI element detection\xspace}  
\newcommand{\ua}{GUI usability and accessibility testing\xspace} 
\newcommand{\eva}{GUI testing evaluation criteria\xspace}

\newcommand{\tabincell}[2]{\begin{tabular}{@{}#1@{}}#2\end{tabular}} 

\setcopyright{acmcopyright}
\copyrightyear{2025}
\acmYear{2025}
\acmDOI{1}

\acmJournal{CSUR}
\acmVolume{1}
\acmNumber{1}
\acmArticle{1}
\acmMonth{1}

\begin{document}

\title{Vision-Based Mobile App GUI Testing: A Survey}

\author{Shengcheng Yu}
\email{yusc@nju.edu.cn}
\orcid{0000-0003-4640-8637}
\affiliation{
\institution{State Key Laboratory for Novel Software Technology, Nanjing University}
\city{Nanjing}\country{China}\postcode{210093}
}
\author{Chunrong Fang}
\authornote{Chunrong Fang is the corresponding author.}
\email{fangchunrong@nju.edu.cn}
\orcid{0000-0002-9930-7111}
\affiliation{
\institution{State Key Laboratory for Novel Software Technology, Nanjing University}
\city{Nanjing}\country{China}\postcode{210093}
}
\author{Ziyuan Tuo}
\email{zytuo@smail.nju.edu.cn}
\orcid{0009-0009-2144-8852}
\affiliation{
\institution{State Key Laboratory for Novel Software Technology, Nanjing University}
\city{Nanjing}\country{China}\postcode{210093}
}
\author{Quanjun Zhang}
\email{quanjun.zhang@smail.nju.edu.cn}
\orcid{0000-0002-2495-3805}
\affiliation{
\institution{State Key Laboratory for Novel Software Technology, Nanjing University}
\city{Nanjing}\country{China}\postcode{210093}
}
\author{Chunyang Chen}
\email{chun-yang.chen@tum.de}
\orcid{0000-0003-2011-9618}
\affiliation{
\institution{Technical University of Munich}
\city{Heilbronn}\country{Germany}\postcode{74076}
}
\author{Zhenyu Chen}
\email{zychen@nju.edu.cn}
\orcid{0000-0002-9592-7022}
\affiliation{
\institution{State Key Laboratory for Novel Software Technology, Nanjing University}
\city{Nanjing}\country{China}\postcode{210093}
}
\author{Zhendong Su}
\email{zhendong.su@inf.ethz.ch}
\orcid{0000-0002-2970-1391}
\affiliation{
\institution{Department of Computer Science, ETH Zurich}
\city{Zurich}\country{Switzerland}\postcode{8092}
}

\begin{abstract}

\hrule\vspace{0.2cm}
Graphical User Interface (GUI) has become one of the most significant parts of mobile applications (apps). It is a direct bridge between mobile apps and end users, which directly affects the end user's experience. Neglecting GUI quality can undermine the value and effectiveness of the entire mobile app solution. Significant research efforts have been devoted to GUI testing, one effective method to ensure mobile app quality. By conducting rigorous GUI testing, developers can ensure that the visual and interactive elements of the mobile apps not only meet functional requirements but also provide a seamless and user-friendly experience. However, traditional solutions, relying on the source code or layout files, have met challenges in both effectiveness and efficiency due to the gap between what is obtained and what app GUI actually presents. Vision-based mobile app GUI testing approaches emerged with the development of computer vision technologies and have achieved promising progress. In this survey paper, we provide a comprehensive investigation of the state-of-the-art techniques on \papernum papers, among which \visnum are vision-based studies. This survey covers different topics of GUI testing, like \gen, \rr, \frm, \etc In particular, we highlight the emerging role of vision-based techniques and analyze how they reshape traditional approaches to mobile app GUI testing. Based on the investigation of existing studies, we outline the challenges and opportunities of (vision-based) mobile app GUI testing and propose promising research directions with the combination of emerging techniques.

\end{abstract}



\maketitle

\section{Introduction}

With the rapid development of mobile applications (apps), the natural GUI-intensive and event-driven features of apps \cite{banerjee2013graphical} make the Graphical User Interface (GUI) of a mobile app hold paramount significance. Mobile devices, with their inherent constraints of screen size and user attention, necessitate interfaces that are both intuitive and efficient. A robust mobile app GUI not only optimizes the user experience but also ensures that complex tasks are executed with minimal friction and cognitive load \cite{linares2017continuous}. As a crucial bridge between end users and apps, a well-designed GUI becomes critical in reducing potential errors in user operations, enhancing user satisfaction, and ultimately determining the success of the app in a competitive marketplace \cite{arnatovich2018systematic}. Within this context, the GUI functions not just as a visual layer but as a crucial mediator between the user intent and the app functionality. Consequently, any single bug in the GUI of mobile apps can have profound implications. A faulty GUI not only jeopardizes the overall user experience but can also lead to potential misbehaviors of the app \cite{liu2020owl}. In some instances, a GUI bug can even expose users to security vulnerabilities, where sensitive data might be at risk \cite{xiao2019iconintent}. Thus, ensuring GUI quality is a significant part of the whole mobile app quality assurance.

The software testing community has been long-term focused on the improvement of mobile app GUI testing effectiveness and efficiency in all aspects since last century \cite{memon1999using}. GUI Testing is an important branch of mobile app testing, the main purpose of which is to verify and ensure that the app GUI conforms to specifications and design requirements by checking whether necessary information is shown in the correct place and whether the test events can lead to intended responses by simulating actual user operations \cite{linares2017continuous, arnatovich2018systematic, said2020gui}. It typically involves testing the functionality, appearance, and interaction of GUI elements (such as buttons, text boxes, menus, \etc) to ensure the normal operation of the user interface and the consistency of the user experience. GUI testing is challenging due to the complex interactions among different GUI components, fragmented running environments, customized GUI widget design, and the faithfulness of user operation simulation. With the development of all kinds of app analysis technologies, GUI testing has gradually evolved from manual testing to automated testing, and automated techniques have taken a prominent place. With the integration of intelligent algorithms, mobile app GUI testing has achieved much progress.

Traditional mobile app GUI testing approaches mainly focus on the source code or layout files of apps under test (AUT) to retrieve information for GUI test generation or execution. However, such traditional approaches are faced with several challenges. First, due to the well-known ``fragmentation problem'' \cite{wei2016taming} of mobile platforms, which means the fragmented running environments of mobile apps, GUI testing is required to be adapted to different environments, which is hard because of completely different implementations or underlying system support. Second, GUI element detection of traditional techniques is to obtain the runtime layout structures from the apps \cite{chen2020object}. However, some GUI elements are missing in the layout files \cite{yu2021layout}. For example, the widely used \texttt{Canvas} widgets are used to customize widget styles and contents. There may exist \texttt{Buttons}, \texttt{TextFields}, \etc, within the \texttt{Canvas} widgets, and such contained widgets will not be revealed in the GUI layout files. Besides, some widgets may constantly refresh their contents, and when the GUI testing techniques try to find the preset targets that have been refreshed, it should be determined whether to relocate the targets by contents or by locations. The inaccurate widget identification further leads to the failure of faithfully simulating user operations.

Briefly speaking, these is a huge gap between the information obtained from the layout files and what is actually presented in the GUI level. According to the mental model \cite{blackwell2006reification}, which is the basis for visual metaphor design in mobile apps, it is a better perspective to start retrieving information in GUI testing from the visual perspective \cite{bajammal2020survey}, which can provide information of ``what you see is what you get''. With the emerging development of computer vision (CV) technologies, the vision-based GUI testing approaches have gradually emerged to solve the aforementioned challenges, which is to analyze and understand the GUI of AUT from the visual perspective (\ie app GUI screenshots). First, app developers typically design highly similar GUI contents and layouts in different running environments, which creates convenience for the universal analysis and understanding of the app GUI. Second, the \ele is enhanced by directly analyzing the app GUI screenshots, similar to human testers. Third, the simulation of user operations can be conducted on the basis of vision-based \ele. 

Mobile app testing is a widely studied topic. Several surveys on mobile app GUI testing emphasize automated testing or the overall evolution of GUI testing \cite{arnatovich2018systematic, banerjee2013graphical, said2020gui}. Several of related studies offer valuable perspectives. For example, a bibliometric overview that charts the growth of GUI testing over time \cite{rodriguez202130}, the survey focused on Android app testing techniques \cite{li2023gui}, or reviews centered around user experience and the human-centric aspects of mobile GUI evaluation \cite{deshmukh2023automated}. Our survey addresses this gap by focusing on source code requirements (\ie black-box, grey-box, white-box) and approach basis (\ie code, layout, image) to trace the evolution of mobile app GUI testing techniques over time. While these surveys offer valuable insights, they differ from our work in scope and focus. Most target general GUI testing, Android-specific issues, or UX considerations, without addressing the broader evolution of mobile GUI testing. Crucially, none examine the emerging role of vision-based techniques or systematically compare them with traditional approaches across core dimensions. Our study is the first, to our knowledge, to provide a comparative and cross-cutting perspective that highlights this shift toward visual reasoning and its implications for future research and tool development. Recent studies have also explored the integration of advanced technologies in software engineering. For instance, Bajammal \etal \cite{bajammal2020survey} examine the application of computer vision (CV) in general software engineering, while others investigate the use of emerging techniques (\eg deep learning) in software development \cite{linares2017continuous, ricca2021ai}. Wang \etal \cite{wang2024software} provide a comprehensive overview of large language models (LLMs) in software testing. However, these surveys lack emphasis on mobile app GUI testing. To address this, our work provides a detailed investigation into the development of mobile app GUI testing, highlighting the transition from non-vision-based to vision-based techniques. Covering eight major topics, we categorize the literature by approach basis and source code requirement. Notably, we present the first comprehensive review focused on \textit{vision-based mobile app GUI testing} techniques.

\begin{table}[!htbp]
\centering
\vspace{-0.3cm}
\caption{Mobile App GUI Testing Research Topics and Approach Taxonomy}
\vspace{-0.3cm}
\scalebox{0.65}{
\begin{tabular}{c|rr|rrr|rrrrrr}

\toprule
\multirow{2}{*}{Topic} & 
\multirow{2}{*}{\# Total Paper} & 
\multirow{2}{*}{\# Vision-based Paper} & 
\multicolumn{3}{c|}{Source Code Requirement} & 
\multicolumn{6}{c}{Approach Basis (C: Code, L: Layout, I: Image)} \\ \cmidrule(l){4-12} 
& &
& \multicolumn{1}{c}{Black-box} 
& \multicolumn{1}{c}{Grey-box} 
& \multicolumn{1}{c|}{White-box}
& \multicolumn{1}{c}{C} 
& \multicolumn{1}{c}{L} 
& \multicolumn{1}{c}{I} 
& \multicolumn{1}{c}{C \& L} 
& \multicolumn{1}{c}{L \& I} 
& \multicolumn{1}{c}{C \& L \& I} \\ \midrule

GEN  & 62  & 10 & 28  & 21 & 13 & 32  & 18 & 10 & 2 & 0 & 0 \\
R\&R & 34  & 13 & 20  & 11 & 3  & 12  & 8  & 12 & 1 & 1 & 0 \\
FRM  & 56  & 8  & 33  & 14 & 9  & 24  & 24 & 8  & 0 & 0 & 0 \\
SRP  & 16  & 3  & 6   & 5  & 5  & 7   & 4  & 2  & 2 & 1 & 0 \\
RPT  & 34  & 14 & 30  & 1  & 3  & 17  & 3  & 14 & 0 & 0 & 0 \\
ELE  & 39  & 36 & 32  & 6  & 1  & 2   & 0  & 31 & 0 & 5 & 1 \\
U\&A & 11  & 6  & 9   & 2  & 0  & 1   & 3  & 5  & 1 & 1 & 0 \\
EVA  & 19  & 2  & 12  & 1  & 6  & 7   & 11 & 1  & 0 & 0 & 0 \\ \midrule
Sum  & 271 & 92 & 170 & 61 & 40 & 102 & 71 & 83 & 6 & 8 & 1 \\ \bottomrule

\end{tabular}}
\label{tab:approach}
\vspace{-0.3cm}
\end{table}

We believe that this survey presents a unique research perspective in the field of vision-based mobile app GUI testing. We, for the first time, focus on vision-based mobile app GUI testing techniques, which have been relatively under-explored in previous secondary research; thus, we consider this in itself to be a novel contribution. Through this unique perspective, we hope to fill the research gap in this field in current reviews and provide a more comprehensive and in-depth understanding of the development of vision-based GUI testing. In this study, we aim to systematically review the landscape of mobile app GUI testing techniques, with a particular emphasis on the emergence of vision-based approaches. Vision-based testing leverages visual elements, such as screenshots and rendered UI states, rather than relying solely on structural or code-based representations. Our goal is to understand how the field of mobile GUI testing has evolved, and how vision-based methods differ from traditional techniques in addressing long-standing problems. To this end, we identify and classify techniques across various testing dimensions, highlighting trends and future directions driven by visual reasoning. While our analysis covers the general landscape of mobile GUI testing, vision-based techniques are not discussed in isolation but are integrated into each aspect of the review. In this survey, we not only introduce the basic concepts and methods of vision-based mobile app GUI testing but also delve into several key topics within this field (see \tabref{tab:approach}\footnote{The full topic names are shown below. We use the abbreviations in the rest of this paper to denote the topics.}), including \gen, \rr, \frm, \srp, \rpt, \ele, \ua, and \eva. The exploration of these topics is based on an in-depth analysis and summary of relevant academic literature, aiming to provide readers with a comprehensive and profound understanding. Furthermore, we highlight the uniqueness and advantages of vision-based GUI testing by comparing it with other non-vision-based methods. We highlight the emerging role of vision-based techniques and analyze how they reshape traditional approaches to mobile app GUI testing. Therefore, we believe that the paper has made substantial contributions to the novelty of vision-based mobile app GUI testing. 

In constructing our topic taxonomy, we adopt a hybrid top-down and bottom-up approach. From a top-down perspective, we refer to SWEBOK V4.0, \textit{Section 6.2: Testing in the Application Domains}, which explicitly discusses GUI testing as a domain requiring specialized techniques and tools. SWEBOK highlights concerns such as event-driven input models, platform-specific interface rendering, and GUI behavior modeling, issues particularly relevant to mobile GUI testing. These points help us outline core testing activities and lifecycle stages that a comprehensive taxonomy should cover, including test generation, execution, and maintenance, as well as supporting infrastructure like frameworks and tools. However, because SWEBOK provides only high-level abstraction and does not offer a detailed classification specifically for mobile GUI testing, we further refine our taxonomy using a bottom-up process. This involves a comprehensive review of our collected primary studies and several relevant secondary surveys. We systematically analyze recurring research topics, technical challenges, and contribution types to identify common themes. This dual approach allows us to construct a taxonomy that is both theoretically grounded and empirically informed, capturing the evolution of the field while maintaining alignment with established software engineering frameworks.

\begin{wrapfigure}{l}{0.6\linewidth}
\centering
\vspace{-0.3cm}
\includegraphics[width=\linewidth]{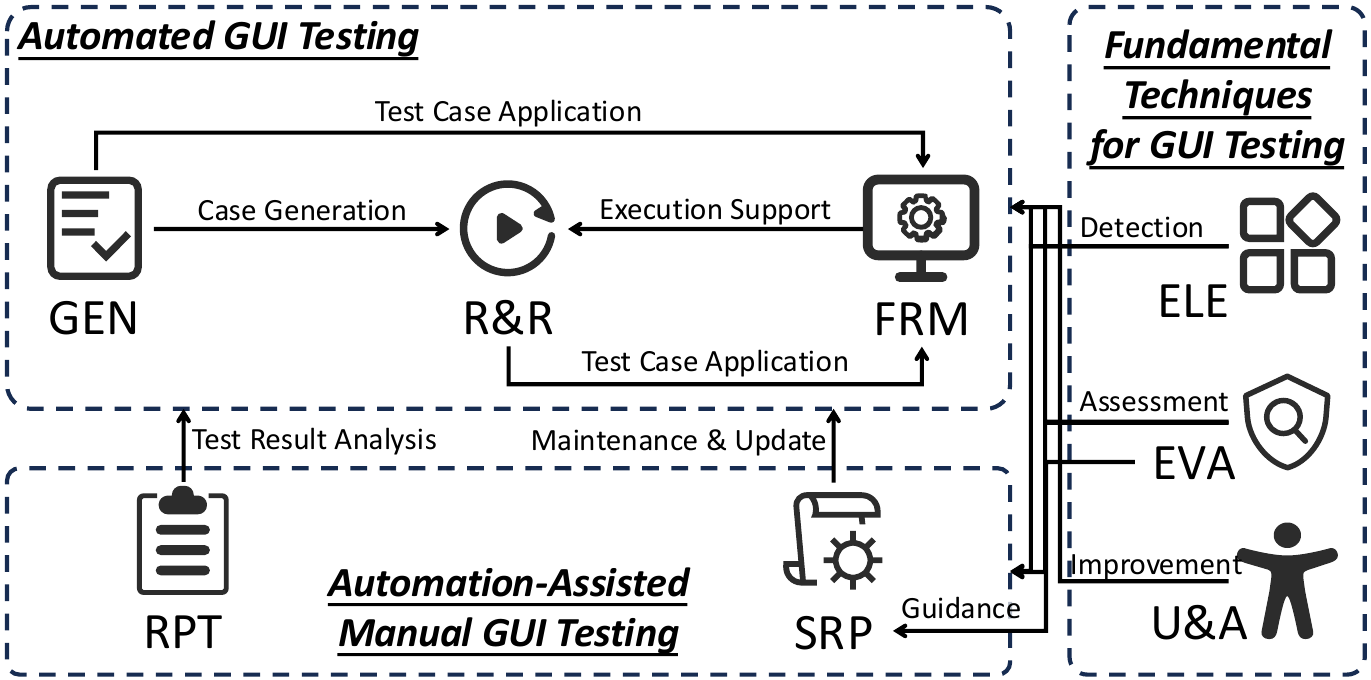}
\vspace{-0.5cm}
\caption{Overview of Mobile App GUI Testing Topics}
\label{fig:overview}
\vspace{-0.3cm}
\end{wrapfigure}

\noindent \textbf{GUI Test Generation (\underline{GEN})}: This topic involves approaches typically designed with a particular optimization goal, such as code coverage, bug detection, \etc, with a specific exploration strategy, like random-based, model-based, and learning-based. It focuses on generating automated test cases that aim to comprehensively explore different GUI behaviors in order to improve overall test coverage and effectiveness.

\noindent \textbf{GUI Test Record \& Replay (Migration) (\underline{R\&R})}: This topic targets at maximizing the value of manually written test scripts, which enables the automated execution of tests on different running environments without manual adaptions. Such test scripts contain the domain knowledge of human testers and can better test the critical functionalities. 

\noindent \textbf{GUI Testing Framework (\underline{FRM})}: This topic provides comprehensive frameworks that enable app developers to independently execute the entire GUI testing process without manually piecing together various testing procedures. Such techniques aim to offer an all-in-one solution for achieving efficient execution of the entire GUI testing process.

\noindent \textbf{GUI Test Script Maintenance (\underline{SRP})}: This topic focuses on the approaches to maintain the test scripts that are not more usable due to the changes in the AUT and to handle the newly developed functionalities of the AUT on the basis of existing test scripts in regression testing.

\noindent \textbf{GUI Test Report Analysis (\underline{RPT})}: This topic deals with the products of GUI testing, the test reports. No matter whether the reports are manually written or automatically generated, the reliability or comprehensibility of the reports needs to be improved, and the quantity of reports should be optimized to help app developers better review these reports.

\noindent \textbf{GUI Element Detection (\underline{ELE})}: This topic specifically focuses on the approaches that deal with the \ele tasks, which is a significant basis of all the vision-based mobile app GUI testing techniques. We discuss traditional CV technologies, advanced learning-based technologies, and their combination.

\noindent \textbf{GUI usability and accessibility testing (\underline{U\&A})}: This topic involves techniques that ensure easy-to-use and accessible GUI, including for those with disabilities. Usability testing focuses on evaluating the effectiveness, efficiency, and satisfaction of users interacting with the interface, ensuring it is intuitive and user-friendly. Accessibility testing, on the other hand, assesses whether the GUI meets the needs of users with varying abilities, including those with visual, auditory, cognitive, or motor impairments.

\noindent \textbf{GUI Testing Evaluation Criteria (\underline{EVA})}: This topic summarizes recent research efforts on evaluating and comparing GUI testing techniques as well as proposing new evaluation standards. The aim is to provide testing professionals with insights that enable the application of these technologies in more suitable and effective testing scenarios.

The relationships of different topics are shown in \figref{fig:overview}. The \gen, \rr, and \frm, which belong to the automated GUI testing, can guide the maintenance and update of \srp, and the results from them can be analyzed through \rpt to provide a better bug inspection reference to app developers. Among the three automated GUI testing topics, \gen and \rr apply the generated test cases or scripts with the support of \frm, and \gen can generate replayable cases for \rr. \ele, \ua, and \eva, which are viewed as fundamental techniques for GUI testing, provide the target detection, specific testing target, and effectiveness assessment, respectively, for both automated GUI testing and automation-assisted manual GUI testing.

\section{GUI Testing Technique Taxonomy}

The primary focus of this paper is on the \textit{vision-based} GUI testing, and how mobile app GUI testing develops from traditional techniques to vision-based ones, so we study on the specific technologies used by the GUI testing studies. We use two taxonomies, the source code requirement and the approach basis (\tabref{tab:approach}). 

For the collected literature, we categorize the paper into black-box, white-box, and gray-box testing, according to source code requirement, as well as methods based on code, layout, image, or a combination, to better reflect the diversity of mobile app GUI testing and current trends in technology development. This classification helps us understand the depth and breadth of testing techniques from different perspectives, particularly the complexities faced in real-world mobile app testing. By categorizing literature, we can clearly illustrate how various testing techniques meet different testing needs. For instance, black-box testing is suitable for validating user experience, while white-box testing is better for analyzing internal logical errors. This classification enables a comprehensive and systematic evaluation of existing research findings and points out potential future research directions. Furthermore, the classification of approach basis on code, layout, and image visually reflects the trends in mobile app testing technologies. Traditional code-based testing methods, while capable of deep analysis, often face limitations due to the complexity and accessibility of code. As the complexity of mobile app interfaces increases, layout- and image-based testing methods have gradually become mainstream, especially in automation and visual testing. Image-based techniques more directly simulate user interactions with apps, providing more meaningful testing scenarios. This classification not only represents the stages of technological development but also highlights the significance of visual testing methods in the mobile app domain.

This classification approach allows us to better understand the overall landscape of mobile app GUI testing technologies, particularly for scenarios requiring automated testing in complex and dynamic environments. It clearly presents the strengths and weaknesses of different technological routes, providing clear guidance for researchers and practitioners. Through this multidimensional classification, we can comprehensively cover various research directions, ensuring that our review contributes representatively and prospectively to the field of mobile app GUI testing. During the paper analysis and classification, three of the authors label the source code requirement and approach basis for all the \papernum papers independently. If the labeling results for the same individual paper is all the same, the labeling is confirmed. If there are inconsistent labeling results, two more senior authors, who have at least 15 years of experience in mobile app GUI testing research and development, are involved in to discuss the labeling results, until a consensus is reached for each inconsistent paper. Further, we conduct the Fleiss kappa test to the results, and the value reaches 0.95, which means ``almost perfect agreement''. This paper analysis process can effectively guarantee the correctness of our paper classification results.

From the statistics, we can observe that on the source code requirement aspect, black-box techniques are utilized at a much higher rate compared to white-box techniques. The reason is that the GUI testing of mobile apps is mostly based on the \textit{.apk} files (Android Package), which do not directly provide the source code. This practice can help protect the privacy and intellectual property of the apps and is more suitable for testing from the integration level. From the approach basis, there are some approaches that utilize the source code in the early years. Besides the studies that directly apply the source code, some may utilize reverse engineering to obtain the source code from \textit{.apk} files to better analyze the AUT. However, in the second stage of mobile app testing, the GUI structure layout files are more commonly used to analyze the apps because of the GUI-intensive and event-driven features of mobile apps. Such layout files are easy to obtain from the \textit{.apk} files and are convenient to use in all topics of GUI testing, but some specially designed GUI widgets are likely to be ignored \cite{yu2021layout, chen2020unblind, yu2024practicalnon}. In recent years, with the development of CV technologies, it has become more and more accurate to identify GUI widgets directly from app GUI screenshots, and some technologies, \eg convolutional neural networks (CNN), optical character recognition (OCR), enable the automated identification of GUI widget attributes, including widget type, attached texts, containing or relative relationships, \etc 
Overall, combined with \figref{fig:year}, we can find that in all directions, the vision-based (\ie image-based) approaches are becoming a strong new force in mobile app GUI testing.

\section{Automated GUI Testing}

Automated GUI testing refers to the approach of using automated tools and scripts to execute GUI tests. This automates the testing process, including tasks such as test generation, interaction execution on the GUI, and checking whether the system's responses align with expected outcomes. This survey investigates three widely studied topics: \gen, \rr, and \frm. A comprehensive exploration of these topics facilitates a better understanding and application of various aspects of automated GUI testing, enhancing testing efficiency, reducing manual errors, and thereby improving software quality and stability.

\subsection{GUI Test Generation}

\begin{figure}[!htbp]
\centering
\vspace{-0.3cm}
\includegraphics[width=0.7\linewidth]{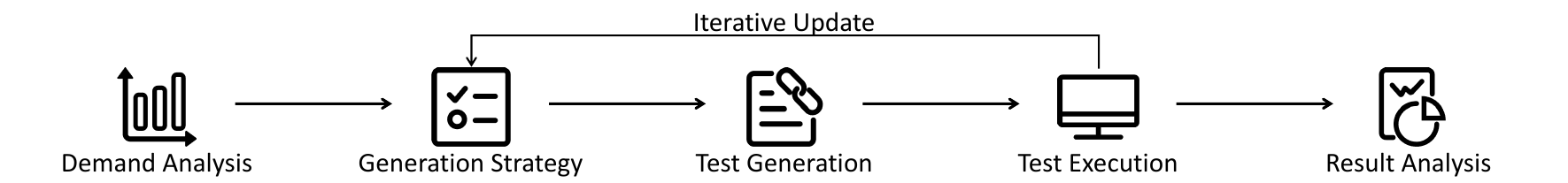}
\vspace{-0.3cm}
\caption{General Workflow of GUI Test Generation}
\vspace{-0.4cm}
\label{fig:gen}
\end{figure}

With the increasing complexity of mobile apps, manual creation of test cases has become tedious and time-consuming. Hence, the emergence of the automated GUI test generation holds significant importance. This part of the survey explores numerous technologies, algorithms, and tools proposed in recent years to automatically generate test cases that effectively encompass diverse GUI interaction scenarios. These auto-generated test cases not only enhance test coverage but also have the capability to uncover potential errors and defects. They provide testing professionals with a more efficient and comprehensive testing approach, aiding in the identification of latent issues and elevating software quality. The general workflow of \gen is shown in \figref{fig:gen}.

\gen includes four main approach clusters, random-based, model-based, system-based, and learning-based approaches (\tabref{tab:gen}). Random-based approaches generate tests by randomly selecting input values or operation sequences. This strategy is often used to reveal unforeseen exceptions or boundary conditions within a system, as well as to conduct rapid and extensive testing of the system. 
Random-based testing approaches initially emerged in the field of hardware system testing, to test the stability and reliability of computer hardware by randomly generating different input signals. This strategy later finds applications in software testing. 
Random-based test generation approaches do not rely on specific models or algorithms, making them suitable for testing software or systems without detailed specifications. Due to the random nature, random testing can help uncover issues that app developers may not have considered, thus aiding in discovering unknown potential problems. Zeng \etal \cite{zeng2016automated} conduct an industry case study on the most typical Monkey \cite{monkey} tool. This work applies random-based GUI test generation to WeChat and reports the limitations of Monkey in an industrial environment. 
However, although random testing can traverse the test path as much as possible, it also causes the problem of redundancy of test cases. The generated test cases lack structure and are not accurate enough.

Model-based strategy is a relatively early group that utilizes abstract models representing the structure, behavior, or interactions of an AUT to generate test cases. The earliest model-based test case generation techniques primarily focus on static and dynamic analysis techniques, such as approaches based on code coverage and path analysis. These approaches can detect some simple errors, but they often cannot generate test cases with high coverage. 
Stoat, introduced by Su \etal \cite{su2017guided}, is used to conduct model-based testing on Android apps. Stoat operates in two phases. 
Model-based approaches rely on specific models or specifications to generate targeted test cases. However, they often suffer from limited coverage and may not explore potential issues extensively.

As mobile apps become more complex, GUI test generation approaches need more system-level considerations. Therefore, system-based test case generation methods are starting to gain widespread use in some large app projects. This strategy includes testing the functionality, performance, and security of the entire app system, focusing on changes and interactions within the entire app system. It can be applied to sensor leakage, crash testing, code changes, \etc 
Navidroid, proposed by Liu \etal \cite{liu2022guided}, constructs a rich state transition graph, plans development paths using dynamic programming algorithms, and enhances GUI with visual prompts, enabling testers to quickly develop untested activities, avoiding redundant development. 
Usually, system-based test case generation relies on the runtime and execution processes of the actual app, capturing real user interactions and behaviors while using the software. This helps discover issues that may arise in real scenarios and can compensate for the limitations of static testing. 
Li \etal \cite{li2017droidbot} propose a GUI-guided test input generator that based on dynamically generated state transition models and allows users to integrate their own strategies or algorithms. It is important to note that test case generation methods based on dynamic event extraction also face challenges such as monitoring and recording a large number of runtime events and determining when to stop testing. Additionally, they may have some impact on system performance due to the need for monitoring and recording runtime events.

With the continuous development of machine learning (ML) and deep learning (DL) technologies and the increasing complexity of app systems, many approaches have had to be completed with the assistance of learning-based methods, and learning-based GUI test generation approaches have begun to rise. This strategy can adjust and improve GUI test generation strategies based on continuously accumulated data. It adapts to app changes and evolution, thereby enhancing the continuous effectiveness of GUI testing. Moreover, with the advancement of modern GUIs, GUI images contain abundant information, including elements such as icons, colors, layouts, and more. Learning-based approaches make vision-based GUI testing possible, overcoming the limitations of code and text analysis. 
Yu \etal \cite{yu2022universally} propose UniRLTest, a reinforcement learning-based GUI test generation method using a general framework including CV algorithms. It extracts GUI widgets from GUI pages and represents the corresponding layout of the GUI. The app is explored guided by a novel designed curiosity-driven strategy. 
Humanoid, proposed by Li \etal \cite{li2019humanoid}, utilizes deep neural network models to understand how human users select actions based on the app GUI from human interaction traces, guiding test input generation for higher coverage. DeepGUI \cite{yazdanibanafshedaragh2021deep} uses a completely black-box and cross-platform approach to collect data, learn from them, and generate heatmaps, thus supporting all scenarios, apps, and platforms. 
Pan \etal \cite{pan2020reinforcement} introduce Q-testing, which employs a curiosity-driven strategy to explore Android apps. This strategy uses a memory set to keep track of previously visited states and guides testing of unfamiliar functionalities. 
However, it is important to note that learning-based GUI test generation approaches come with challenges, such as the need for large-scale training data and the selection of suitable learning models and algorithms. Moreover, these methods may not be suitable for all types of apps and testing scenarios, so their applicability needs to be assessed based on specific circumstances.

The development sequence of these categories may be influenced by various factors, including technological advancements, changing requirements, \etc Therefore, it is hard to say that one strategy is more advanced than another. In practical applications, the choice of a suitable GUI test generation approach is often based on specific testing requirements and technological feasibility. Additionally, these methods may be combined to achieve better test coverage and effectiveness.

Meanwhile, it is necessary to explore the fact that, despite the convenience provided by various automated GUI testing generation technologies for testing across multiple devices and apps, and their excellent performance in compensating for the shortcomings of manual testing, there are still efficiency issues in existing GUI testing generation technologies, particularly in the aspect of UI exploration. 
As GUI serves as the primary means of interaction between users and apps, its efficiency directly impacts user experience and the overall performance of the app. The key to solving this problem lies in developing smarter and more efficient UI exploration strategies to ensure that as many app functions as possible are covered in a reasonable time, thereby improving GUI efficiency and overall testing effects. 
Wang \etal \cite{wang2021vet} propose VET, a general method for effectively identifying and resolving exploration tarpits issues, which can be automatically applied to enhance any Android UI testing tool on any AUT.
Feng \etal \cite{feng2023efficiency} propose a method based on images that take into account the real-time stream on the GUI, provide a deep learning model to infer rendering status, and synchronize with the testing tool to schedule the next event when the GUI is fully rendered.

In addition, factors such as GUI response time, GUI resource usage, GUI design, GUI implementation, GUI testing techniques, and the hardware and software of mobile devices can all have an impact on GUI efficiency. Corresponding optimization strategies can be adopted for these different influencing factors. Therefore, GUI efficiency is an important and complex issue in the GUI testing of mobile apps. When addressing this issue, it is necessary to comprehensively consider various evaluation indicators, influencing factors, and optimization strategies, and to verify and analyze test results through experiments. Such comprehensive consideration helps to formulate a comprehensive and effective GUI testing strategy to ensure that the app achieves the best performance level in terms of the GUI.

\subsection{GUI Test Record \& Replay (Migration)}

\begin{figure}[!htbp]
\centering
\vspace{-0.3cm}
\includegraphics[width=0.8\linewidth]{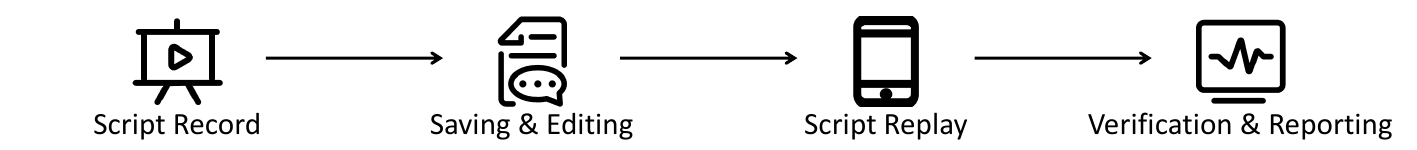}
\vspace{-0.3cm}
\caption{General Workflow of GUI Test Record \& Replay (Migration)}
\vspace{-0.3cm}
\label{fig:rr}
\end{figure}

In the field of GUI testing, GUI test record and replay is becoming increasingly prominent. This part of the survey examines various recording and replaying approaches, which share the common focus of capturing user interactions with the app GUI, recording them as test scripts, and subsequently replaying these interactions to verify the software's behavior in different scenarios. The emergence of such technologies has significantly simplified the task of test script creation for testing professionals, reducing skill barriers and enabling non-technical individuals to participate in GUI testing. The general workflow of \rr is shown in \figref{fig:rr}.

Initially, the main purpose of \rr is for the same app on the same platform, for regression testing. This category always focuses on similar external environments, ignoring the mobile app fragmentation problem. 
Such approaches are studied to reduce the number of recorded events, replay error reports using image processing techniques, and combine GUI exploration with record \& replay technologies. 
With the development of mobile devices and platforms, the fragmentation problem \cite{wei2016taming} requires \rr techniques to conduct cross-device or even cross-platform migration (\tabref{tab:rr}). Cross-device approaches focus on different types of devices (such as smartphones, tablets, desktop computers), different brands of devices (like Huawei, Samsung, Xiaomi), or different versions of operating system devices (like Android 7, Android 9, Android 10) within the same operating system environment. These approaches allow app developers to create scripts on different devices of the same operating system and replay them on those devices to ensure the proper functionality of the apps on various devices. Cross-platform approaches encompass technologies that enable record \& replay across different operating system platforms, such as Android and iOS, Android and Web, \etc 
This means that this technology allows the creation of test scripts on different operating systems and their replay on various platforms to validate the consistency and compatibility of the apps across different platforms. 
Then, with the emergence of a large number of mobile apps, people hope to migrate tests to different apps but with similar functionalities. 
Cross-app approaches indicate technologies that enable record \& replay between different apps. 
It can be based on similar scenarios within different apps. This means that this technology can capture and simulate user interactions across different apps, often used to validate common scenario-based tests across various apps.

Early on, mobile app GUI testing primarily relies on manual testing, but some basic record \& replay tools begin to emerge for capturing and replaying user interactions on mobile apps. These tools typically focus on simulating basic user actions like clicking, swiping, and inputting to validate the app core functionality. jRapture \cite{steven2000jrapture} captures interactions between Java programs and systems, 
ensuring that during replay, each thread sees the exact same input sequence captured during the record. 
Early technologies have limitations and are mostly suitable for validating basic app functionality. Automated testing frameworks provide more robust capabilities, allowing testers to script complex automation test scenarios. 
Record \& replay functionality gradually integrates deeply with some automated testing frameworks, enabling app developers to achieve advanced testing scenarios and custom operations through scripts. 

The diversity of mobile devices continues to grow, and record \& replay approaches begin to support record \& replay GUI tests on different types of devices, such as smartphones and tablets, and different versions of operating systems, device models, hardware features, \etc This ensures compatibility of the apps across various devices and OS versions to meet the needs of a broad user base. 
As an exploratory work, Li \etal \cite{li2022cross} demonstrate that leveraging the Minimum Surprise Principle in GUI design and using a greedy algorithm can simplify and make cross-device record \& replay practice. 
Simultaneously, as the mobile app ecosystem evolves and cross-platform apps gain popularity, record \& replay approaches begin to support app GUI testing on different operating system platforms like iOS and Android. This enables app developers to use similar tools for record \& replay GUI tests on different platforms. 
TestMig \cite{qin2019testmig} is proposed for migrating iOS GUI test cases to Android. Specifically, it records the sequence of events invoked during iOS testing and uses sequence transformation techniques to convert this sequence into an Android event sequence.

In discussing cross-device and cross-platform record-and-playback technologies, it is evident that in addition to addressing the challenges of compatibility across platforms and devices, it is crucial to consider the substantial costs associated with generating test cases for app development, especially pertinent when dealing with a multitude of diverse apps. Consequently, some studies suggest that by taking into account the similarities between apps and reusing test cases among similar apps, the cost of testing mobile apps can be significantly reduced. Behrang \etal \cite{behrang2018automated} propose AppTestMigrator, an approach that leverages the similarities between the GUIs of different but related apps to reuse and adapt test cases, thereby reducing the cost of app testing.

In addition, since the recording and replaying technology is usually implemented through black box technology, it is more inclined to observe and simulate the appearance and behavior of the user interface, so it is more dependent on the visual information of the GUI. Through visual analysis, the interaction between users and apps can be captured, including button clicks, text input, and the display and hiding of interface elements. 
In recent years, machine learning and deep learning techniques have begun to be applied to mobile app test recording and playback to help automatically generate test scripts, optimize test cases, and detect potential problems.
These technologies can enhance the effectiveness and efficiency of GUI testing, reduce testing workload, and extract information from a visual perspective as supplementary text information. An image-driven cross-platform test script record \& replay approach, LIRAT \cite{yu2021layout}, is proposed, addressing cross-platform record \& replay for test scripts. 
V2S \cite{bernal2020translating} is introduced primarily through CV techniques like object detection and image classification, effectively identifying and classifying user behaviors in screen record video frames, thereby converting them into replayable test scenarios. 
GIFdroid \cite{feng2022gifdroid} is a lightweight approach that automatically replays execution traces from visual error reports. 

The development of mobile app GUI test record \& replay technology also reflects the progress of mobile technology and evolving testing requirements. These advancements help improve the quality, performance, and user experience of mobile apps, providing developers and testers with more tools and resources to ensure correct app behavior in various situations. As mobile technology continues to evolve, these record \& replay approaches continue to evolve to meet new challenges and opportunities.

\subsection{GUI Testing Framework}

\begin{figure}[!htbp]
\centering
\vspace{-0.3cm}
\includegraphics[width=0.8\linewidth]{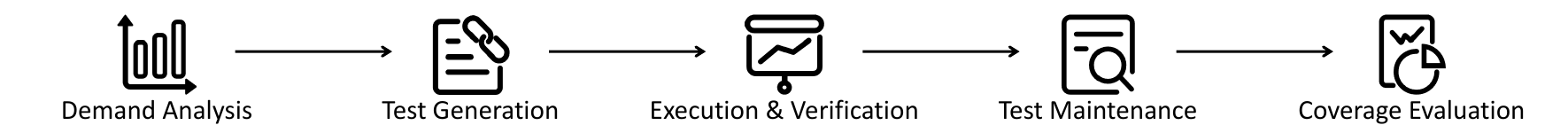}
\vspace{-0.3cm}
\caption{General Workflow of GUI Testing Framework}
\vspace{-0.3cm}
\label{fig:frm}
\end{figure}

In the evolution of GUI testing, while individual research efforts dedicated to specific modules such as test script debugging, test generation, and expected outcome assessment are undoubtedly important, in practical applications, testing professionals increasingly desire a comprehensive framework that enables them to independently execute the entire GUI testing process without manually piecing together various testing procedures. Consequently, this section on testing frameworks compiles the classic frameworks proposed in recent years, aiming to offer testing personnel an all-in-one solution for achieving efficient execution of the entire GUI testing process. The general workflow of \frm is shown in \figref{fig:frm}. While the topics of GUI test generation and GUI testing framework may overlap, they differ in functionality and scope. For studies classified in the GUI testing framework topic, a comprehensive testing environment is presented that provides all necessary tools and components for executing, managing, and reporting tests. It encompasses various aspects, including test execution, management, and result analysis, in addition to test generation. For studies classified in the GUI test generation topic, as an important sub-function within the testing framework, focus specifically on how to automatically generate effective tests. While it is a part of the GUI testing frameworks, it does not include the execution and management functionalities of tests. This distinction highlights the broader role of GUI testing frameworks compared to the more focused nature of GUI test generation.

GUI testing frameworks are designed to automate and simplify the GUI testing process of mobile apps. They allow testers to create, execute, and manage test cases to verify that the app GUI functions as expected, including detecting potential issues and ensuring the stability, functionality, and performance of the app. GUI testing frameworks typically provide various features such as recording and replaying, script automation, cross-platform, and cross-browser support to help development teams ensure a high-quality user experience.

From a technical perspective, there are several types of testing frameworks based on traditional technologies (\tabref{tab:frm}). Module or library-based testing frameworks that encapsulate operations on various modules of the tested product or use a library structure to implement cross-module business operations. Data-driven testing frameworks separate data from test cases, enabling rapid test case generation. Keyword-driven frameworks involve writing test cases in tabular or other formats and converting specific keyword text into executable functions. Behavior-driven development frameworks focus on developing and testing based on explicit expected behaviors.

Mu \etal \cite{mu2009design} propose an extensible, reusable, and easily maintainable GUI automation testing framework that uses XML to store data. It combines data-driven and keyword-driven approaches, separating scripts, data, and business logic to enhance reusability.
In terms of performance optimization, many frameworks have different areas of focus. Some focus on redundant sequence identification, shortening sequence lengths, and sequence generation. For instance, Cheng \etal \cite{cheng2017systematic} propose a new GUI testing framework for effectively generating event sequences while avoiding redundancy. 

However, as mobile apps become increasingly complex, especially those involving large-scale distributed systems and emerging technologies, traditional testing methods and tools may become less flexible and efficient, especially with the richness of visual content in apps. Innovative technologies can better address the complexity of visual phenomena, providing more accurate testing and better issue detection.

Early research into visual issues included Sikuli, proposed by Yeh \etal \cite{yeh2009sikuli}, which allows users to take screenshots of GUI elements and use these screenshots instead of element names to query a help system. Sikuli also provides a visual script API to automate GUI interactions, guiding mouse and keyboard events using screenshot patterns. 


Some research is based on deep learning. Qian \etal \cite{qian2020roscript} introduce RoScript, which employs non-invasive CV technology to automatically record touchscreen actions from human action videos on test devices into test scripts. 
In addition, some novel approaches have also driven the development of GUI testing frameworks. Nguyen \etal \cite{nguyen2014guitar} design a framework that supports the development of customized GUI testing tools. 

In summary, the evolution from early testing frameworks based on traditional technologies to frameworks based on innovative technologies aims to address the challenges of modern mobile app development and delivery, improving testing efficiency, accuracy, and adaptability. This evolution reflects technological and market trends and aims to meet the increasingly complex and diverse app testing requirements.

\section{Automation-Assisted Manual GUI Testing}

Automation-assisted manual GUI testing refers to the approach of utilizing automated techniques to assist in manual user interactions during GUI testing, often applied to complex GUI and scenarios requiring human intervention, and the corresponding test result processing. This combines the strengths of both automation and manual testing, leading to improved testing efficiency and accuracy. This paper emphasizes the study of two major topics within automation-assisted manual GUI testing: \srp and \rpt. These topics focus on ensuring the timely adaptation and maintenance of test scripts for evolving interfaces and functions, as well as on analyzing test results through automated means to extract key insights, providing more accurate decision support and feedback for app developers. 

\subsection{GUI Test Script Maintenance}

\begin{figure}[!htbp]
\centering
\vspace{-0.3cm}
\includegraphics[width=0.8\linewidth]{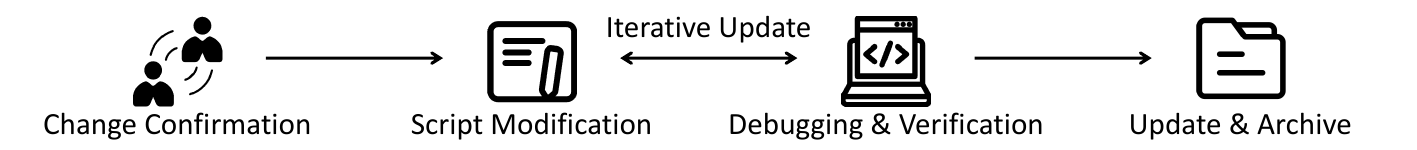}
\vspace{-0.3cm}
\caption{General Workflow of GUI Test Script Maintenance}
\vspace{-0.3cm}
\label{fig:srp}
\end{figure}

As apps continue to evolve and update, test scripts need corresponding adjustments and updates to remain in synchronization with the app. This part explores various technologies and tools proposed in recent years to help testing professionals manage and maintain test scripts more efficiently. For instance, techniques such as automated identification of affected segments within test scripts, methods for automatically updating scripts to adapt to changes, and ensuring that updated scripts still accurately test the apps have been studied. This series of techniques equip testing professionals with means to address changes, thus ensuring the continuous effectiveness of testing. The general workflow of \srp is shown in \figref{fig:srp}.

GUI testing scripts provide essential tools for automated testing, significantly enhancing testing efficiency, repeatability, and coverage. By writing and executing scripts, testing teams can swiftly and accurately uncover potential software defects, ensuring the quality and stability of apps. Regular script maintenance and updates also guarantee compatibility with the latest app versions, playing a crucial role throughout the software development cycle. Therefore, GUI testing scripts are indispensable for ensuring high-quality software delivery.

The script generation phase focuses on automatically creating test scripts from testing requirements and specifications. This involves translating test cases, test steps, or other testing requirements into executable code, ensuring that the system performs expected test operations during automated testing, thus improving testing efficiency and repeatability. The primary goal of GUI test generation is to create test cases that cover various usage scenarios and functionalities to ensure correctness and robustness, including code analysis, GUI element localization, and user operation logging. The core of this process lies in automating the generation of different types of tests, thereby improving test coverage and efficiency. In contrast, test script generation focuses on creating scripts designed to execute specific testing tasks. Unlike GUI test generation, test script generation is more concerned with how to automate the creation of scripts that can perform complex testing steps, allowing these steps to be reused in different environments. Test script generation not only involves the creation of test cases but also requires managing the sequence and logic of test execution to ensure that the testing steps are executed correctly.

Chang \etal \cite{chang2010gui} introduce a novel approach that utilizes CV to achieve GUI testing, extracting component images interacting with the system and visual feedback seen by the user, and automatically generating visual test scripts. Additionally, Iyama \etal \cite{iyama2018automatically} present the time-consuming nature of script generation by proposing a method that automatically generates test scripts through static and dynamic analysis of the app source code and executable files. Yandrapally \etal \cite{yandrapally2014robust} generate test scripts to locate GUI elements by encoding contextual clues. The scripts are independent of tools or platforms and can adapt well to changes in internal attributes or visual rendering.

As mobile apps continue to evolve, test scripts also require regular updates to adapt to new features, requirements, and interfaces. The goal of script updates is to ensure that the test suite remains aligned with the latest software system version, facilitating the timely detection and reporting of any new issues or errors. For instance, Li \etal \cite{li2017atom} introduce ATOM, which utilizes Event Sequence Models (ESMs) to abstract possible event sequences on the GUI and uses Delta ESMs to abstract changes made to the GUI. Given these two models as input.

Test scripts are human-centric. Script refactoring focuses on optimizing existing test scripts to enhance their readability, maintainability, and performance. Through refactoring, scripts become clearer, reducing redundant code and improving script maintainability, thus lowering the cost of test maintenance. Chen \etal \cite{chen2008gui} propose an ``object-based'' approach called component abstraction to model the structure of the GUI. 

During the GUI testing process, scripts may become ineffective due to system changes, errors, or other reasons. The task of script repair is to promptly detect and rectify these issues to ensure the continued reliability of testing. 
Gao \etal \cite{gao2015sitar} introduce a novel script repair approach called SITAR, which utilizes reverse engineering techniques to create abstract tests for each script.
Xu \etal \cite{xu2021guider} propose a mobile app GUI test script repair method called GUIDER, which leverages the structural and visual information of widgets on the app GUI to better understand changes in widgets between the base and updated versions.

In conclusion, script generation, script refactoring, and script updates contribute to improving testing quality and reducing maintenance costs, while script repair ensures the ongoing reliability of testing, adapting to continuously evolving software systems. All of these efforts collectively ensure successful software testing and high-quality delivery, playing a critical role throughout the entire software development cycle.

\subsection{GUI Test Report Analysis}
\label{sec:rpt}

In the complete testing process, test reports serve as the final phase, providing testing professionals with detailed data about test coverage, errors, \etc Testers need to base their subsequent actions on the content of the report. Therefore, analysis and understanding of the report are of utmost importance. This part examines research on test report analysis in recent years, encompassing techniques such as deduplicating reports, prioritizing report hierarchies, and comprehending visual and textual information within reports. These advancements enable testing practitioners to swiftly capture critical information from test reports, thus reducing significant manual effort.

Through test report analysis, the testing team and stakeholders responsible for software quality gain profound insights into the app status, including identified defects, test coverage levels, and the number of passed test cases. This process aids in the early detection of potential issues, guides decision-making, optimizes testing strategies, and provides strong support for app improvements, ultimately ensuring the delivery of high-quality apps.

Crowdsourced testing is a method of conducting software testing using external resources \cite{yu2021prioritize}, typically distributed to individuals or groups. It harnesses the wisdom and power of the crowd by gathering extensive user feedback and operational data to identify and fix defects in software \cite{yu2019crowdsourced}. Compared to traditional internal testing methods, crowdsourced testing offers advantages such as broader coverage, lower costs, and diverse testing environments, making it especially suitable for software like mobile apps that have extensive user bases \cite{cao2020stifa}. In GUI testing, crowdsourced testing enables a more comprehensive coverage of various usage scenarios and operational paths by collecting feedback from users across different devices and environments. Crowdsourced testing is an important branch of GUI testing \cite{yu2023mobile}, providing comprehensive and efficient testing solutions through widespread user participation and diverse testing environments. Despite facing challenges, with the right tools and methods, crowdsourced testing holds significant potential for improving software quality and user experience. However, report processing is a significant issue in crowdsourced testing in mobile app GUI testing. In this part, we have an emphasis on the reports from crowdsourced testing, but also pay attention of GUI test reports in other forms and from other resources.

Test report analysis is a crucial phase in the GUI testing process, involving detailed examination and interpretation of test results, error information, coverage data, and other relevant metrics. Before the rise of the application of CV technology, the analysis of test reports by testers mostly stayed at the text level. For example, 
Chen \etal \cite{chen2021effective} propose a model called RCSE, which encodes the description information of test reports at the sentence level, and calculates the similarity between test reports according to the feature similarity of the described sentences. 

Report duplication detection focuses on identifying and merging similar or duplicate issue reports within test reports. This helps avoid redundant work, improves efficiency, and ensures that each issue is recorded only once, reducing confusion and redundancy. Wang \etal \cite{wang2019images} introduce SETU, which extracts four features from screenshots and textual descriptions and designs a hierarchical duplicate detection algorithm based on four similarity scores obtained from these features. Cooper \etal \cite{cooper2021takes} present TANGO, which employs custom CV and text retrieval techniques to analyze visual and textual information in mobile screen recordings, aiming to generate a list of candidate videos similar to the target video report. Duplicate reports, when identified, can be marked as duplicates in the issue tracker. 

Report clustering involves grouping issues within test reports based on their similarity or relevance. Clustering enables a better understanding of patterns and trends in issues, helping testing teams focus on resolving similar types of problems, thereby improving issue resolution efficiency. 
Liu \etal \cite{liu2020clustering} utilize image understanding techniques, employing spatial pyramid matching to measure the similarity of screenshots and natural language processing (NLP) to calculate textual distances in test reports. They perform clustering of crowdsourced test reports based on the obtained text and image features. 

In cases where a large number of issue reports exist, report prioritization becomes a critical task. It involves ranking issue reports based on factors such as severity, impact, and urgency to ensure that the most critical issues are addressed first, thereby minimizing potential risks. Feng \etal \cite{feng2015test, feng2016multi} present the use of image understanding techniques to assist in determining test report priorities. This method combines a mixture of textual descriptions, screenshot images, and two sources of information to create a hybrid ranking approach. Yu \etal \cite{yu2021prioritize} introduce DeepPrior, which considers a deep understanding of app screenshots and textual descriptions. It transforms app screenshots and textual descriptions into four different features. These features are then aggregated, and DeepSimilarity is calculated based on these features. Finally, report prioritization is carried out according to predefined rules.

Additionally, this field encompasses a range of innovative research directions. Some studies focus on using dialogue systems to assist in report generation, employing NLP and artificial intelligence technologies to automatically generate and explain test reports. For example, Song \etal \cite{song2023burt} introduce BURT, a web-based chatbot for interactive reporting of vulnerabilities in mobile apps. Some researchers work on report enhancement, enriching report content by adding additional contextual information, screenshots, or screen recordings. Fazzini \etal \cite{fazzini2022enhancing} propose EBUG, a method for enhancing mobile app error reports. 
Some researchers focus on report crash reproduction, analyzing and reproducing crash issues to assist developers in fixing them. ReCDroid(+), introduced by Zhao \etal \cite{zhao2019recdroid, zhao2022recdroid+}, combines NLP and dynamic GUI exploration to synthesize event sequences with the goal of reproducing crash reports. 
There are also various other dimensions of innovative research, all aimed at improving the quality, understandability, and practicality of test reports.

The aforementioned studies collectively constitute the diversity of test report analysis, and these categories are not entirely independent; they often complement each other based on testing requirements. In summary, effective analysis of test reports contributes to enhancing the efficiency and quality of the software testing process, ensuring the delivery of high-quality software products.

\section{Fundamental Technique for GUI Testing}

Fundamental techniques for GUI testing refer to various approaches that provide support and foundation for GUI testing. These techniques include GUI image recognition and testing evaluation, and more. This survey primarily investigates two topics: \ele and \eva within the context of fundamental techniques for GUI testing. Among these, image recognition technology can assist automated tools in accurately identifying and interacting with elements on the app GUI during GUI testing. Testing evaluation can help app developers more efficiently select appropriate testing techniques. These auxiliary methods have the potential to enhance the efficiency and accuracy of GUI testing, reducing the workload for app developers.

\subsection{GUI Element Detection}

In the process of creating, auto-generating, and executing test scripts, the significance of GUI element detection and localization cannot be overlooked. This paper recognizes that in recent years, numerous GUI element detection tools and algorithms have been introduced, focusing on how to accurately identify and locate various GUI elements such as buttons, text fields, and drop-down menus, through technologies like image recognition and positioning algorithms. These tools and algorithms play a pivotal role in supporting tasks like test generation and record and replay, aiding testing professionals in effectively managing various types of mobile apps.

With the proliferation of a large number of mobile apps, the visual information contained within mobile apps has become increasingly rich and integral to the testing process. Testers often need to perform GUI element detection to gather information effectively. 
In GUI test record and replay, the recording process often needs to record the GUI elements involved in user operations, and the replay process needs to complete the matching work of replay interface elements and recording interface elements. 
V2S, proposed by Bernal-Cardenas \etal\cite{havranek2021v2s}, is based on CV technology, using object detection and image classification solutions to detect and classify user actions in videos and convert them into replayable tests. 
As an auxiliary technique in GUI testing, the field of GUI element detection has seen a wealth of research output.

\begin{figure}[!htbp]
\centering
\vspace{-0.3cm}
\includegraphics[width=0.8\linewidth]{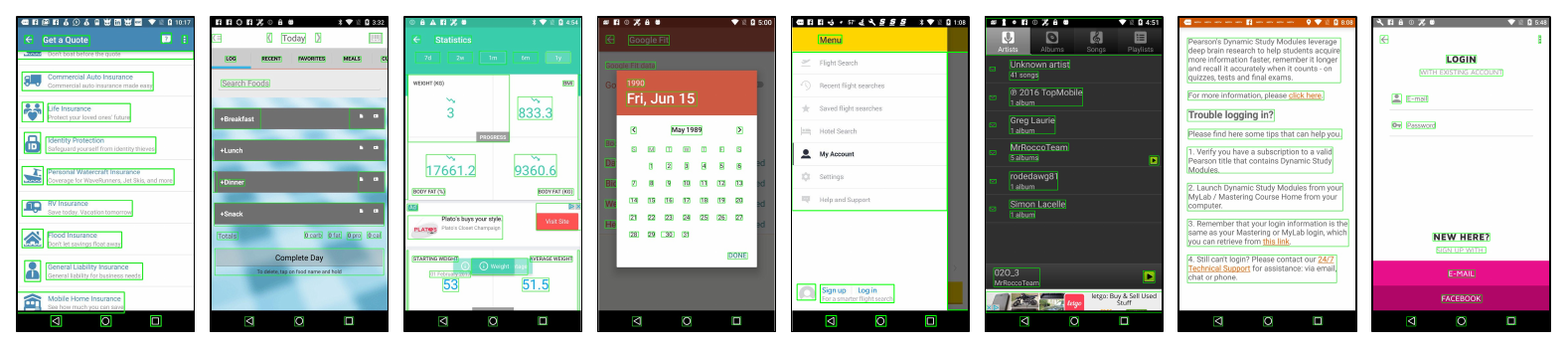}
\vspace{-0.5cm}
\caption{Element Detection with the Algorithm in \cite{chen2020object} and \cite{yu2021layout}}
\label{fig:ele}
\vspace{-0.5cm}
\end{figure}

In recent years, approaches based on image understanding for recognizing app GUI components have provided new solutions and methods for scenarios where automatic component localization is challenging. Some studies focus on GUI element classification. Sun \etal \cite{sun2020ui} employ image understanding to analyze and extract component images from screenshots, design and implement convolutional neural networks (CNNs), and use pre-trained CNNs to classify these images. 
OwlEye, proposed by Liu \etal \cite{liu2020owl}, models the visual information of GUI screen screenshots, enabling the detection of display issues within GUIs and pinpointing the specific areas in the given GUI that need developer attention. 
Furthermore, the iterative updates of GUIs have increased the complexity of testing. Moran \etal \cite{moran2018detecting} introduce an automated method that utilizes CV techniques and natural language generation to accurately and concisely identify changes made to mobile app GUIs between consecutive submissions or releases. Chen \etal \cite{chen2020object} propose a novel GUI element detection method (\figref{fig:ele}).

GUI element detection has also seen innovative research applied to specialized mobile app scenarios or the detection of widgets with unique attributes. 
For instance, due to the highly dynamic and frequent updates in gaming apps, traditional mobile testing techniques may not be directly applicable. 
Ye \etal \cite{ye2021empirical} conduct a survey of the pain points and challenges of game testing and integrate state-of-the-art methods for mobile GUI widget detection, creating a benchmark specifically designed for researching mobile game GUI widget detection. 
In the case of apps used in specialized scenarios, such as aerospace control app, which has stringent quality requirements due to potential safety implications, Zhu \etal \cite{zhu2021gui} introduce a widget region detection technique based on image understanding and analysis. 
Thus, identifying suspicious sensitive data and understanding the intentions behind GUI widgets are crucial. Xiao \etal \cite{xiao2019iconintent} propose a novel framework that utilizes program analysis techniques to associate icons with GUI widgets and employs CV techniques to categorize the associated icons into eight sensitive categories. 

Moreover, there are some related but indirect techniques that provide the foundation for GUI testing, such as GUI prototype generation. This can be used for early visual assessment, user interface feedback, test case generation, layout, and style consistency checks, among other purposes. 
Zhao \etal \cite{zhao2021guigan} develop a GUI design automatic generation model, GUIGAN, which represents the first study in automatically generating mobile app GUI designs. 

In summary, GUI element detection plays a pivotal role in GUI testing, providing testers with powerful tools for identifying, locating, and understanding GUI elements. These studies have not only improved the efficiency and accuracy of GUI testing but have also expanded their applicability to address the testing needs of special scenarios, potentially providing robust support for a wider range of mobile app testing tasks.

\subsection{GUI Usability and Accessibility Testing}

The usability and accessibility of the GUI are crucial aspects in the GUI testing of mobile apps, as they directly affect the user’s experience and satisfaction. In this survey, we take the testing on usability and accessibility as a independent topic to indicate the importance. Usability involves the ease of use and intuitiveness in the user’s interaction process with the app, while accessibility refers to the extent to which the app adapts to different user abilities and preferences, such as visual, auditory, or motor impairments. Overcoming these impairments often relies on GUI element detection technologies.

Currently, the challenge of testing GUI usability often lies in considering the diversity and variability of mobile devices and environments. Mobile devices have different screen sizes, resolutions, orientations, input methods, and hardware capabilities, all of which may affect the layout and performance of the GUI. Moreover, mobile users may use apps in different environments, such as indoors or outdoors, noisy or quiet, bright or dark, stable or moving, \etc These factors can all affect the user’s perception and expectations of the app GUI. Existing GUI testing techniques focus on simulating user interaction with the GUI to trigger app functions. Their effectiveness is evaluated through code and GUI coverage, often unable to verify the visual effects of GUI design. Zhao \etal \cite{zhao2020seenomaly} are the first to check GUI animation effects based on design guidelines, proposing an unsupervised, CV-based adversarial auto-encoder to solve linting problems. In addition, due to software or hardware compatibility issues, display problems such as text overlap, screen blur, and image loss often occur when rendering the GUI on different devices. These can have a negative impact on the usability of the app, leading to a poor user experience. To detect these problems, Liu \etal \cite{liu2020owl} propose a deep learning-based method for modeling the visual information of GUI screenshots to detect display issues and guide developers in repairing bugs.

Testing GUI accessibility also faces a series of challenges, as it requires a comprehensive consideration of the needs and preferences of users with different abilities and impairments. For example, some users may rely on screen readers or voice commands to interact with apps, while others may need larger fonts or higher contrast to clearly see the app. Moreover, some users may have temporary or situational impairments that affect their interaction with the app, such as low battery power, poor network, or background noise. There are already many related studies dedicated to overcoming these challenges. Su \etal \cite{su2022metamorphosis} propose an automated method called dVermin, which detects scaling issues by detecting inconsistencies in views under default and larger display ratios. Chen \etal \cite{chen2020unblind} develop a deep learning-based model that automatically predicts the labels of image-based buttons by learning from large commercial apps. Alshayban \etal \cite{alshayban2022accessitext} propose an automated testing technique called AccessiText to address text accessibility issues when using text scaling auxiliary services. 

\subsection{GUI Testing Evaluation Criteria}

With the emergence of various technologies, frameworks, and tools in GUI testing, questions arise regarding the effectiveness of their testing outcomes, the specific aspects they emphasize in testing, and how testing professionals should select among these technologies. This section of the paper summarizes recent research efforts that have focused on evaluating and comparing these technologies, as well as proposing new coverage criteria or evaluation standards. The aim is to provide testing professionals with insights that enable the application of these technologies in more suitable and effective testing scenarios.

Research on the assessment and establishment of standards for GUI testing provides structured principles that enable testing teams to plan, execute, and evaluate GUI tests with greater rigor and consistency. Such frameworks minimize subjectivity, enhance software quality, and foster standardization across projects and organizations. As the field expands and numerous testing tools emerge, systematic evaluation becomes essential for selecting appropriate methods and guiding future research. Amalfitano \etal \cite{amalfitano2015conceptual} propose a conceptual framework for comparing GUI testing techniques, while Su \etal \cite{su2021benchmarking} present an empirical study evaluating Android GUI testing tools from a complementary perspective.

The evaluation of testing tools and the definition of corresponding assessment standards are critical components of GUI testing research. Establishing standardized evaluation metrics ensures that testing quality and performance are measurable, comparable, and controllable, thereby improving the consistency and reliability of test outcomes. Memon \etal \cite{memon2001coverage} decompose GUIs into components as the fundamental testing units, and Baek \etal \cite{baek2016automated} propose a multi-level comparison standard for more precise GUI evaluation and design a corresponding testing framework.

Furthermore, GUI testing evaluation encompasses broader dimensions, including industrial assessments and user experience evaluations, which provide comprehensive insights beyond functional verification. Such studies enhance both software quality and user satisfaction while advancing GUI testing methodologies to meet the demands of modern applications. Xie \etal \cite{xie2006studying} empirically examine the influence of test case characteristics—such as number, length, and event composition—on cost and fault detection effectiveness.

Collectively, these studies offer testing teams a foundation for informed method selection, ensuring measurable quality and providing valuable guidance for continued innovation in GUI testing research and practice.
\section{Challenges \& Opportunities of Vision-based GUI Testing}

In this section, we delve into the challenges and opportunities of vision-based GUI testing. Such challenges and opportunities widely affect different topics mentioned above, which we will illustrate in the following subsections in detail. Our hope is that this part has provided a deeper understanding of the current challenges in vision-based GUI testing, so as to inspire new opportunities to address these challenges and further improve testing effectiveness and efficiency.

\subsection{GUI Element Detection and Semantics Understanding}

GUI element recognition, \ie GUI widget recognition, is a significant foundation of vision-based GUI testing. It refers to the accurate localization and operation of various elements on the GUI interface, such as buttons, menus, text boxes, lists, \etc, by some technical means. Currently, there are still some challenges in accurate GUI widget recognition in mobile app GUI testing research.

GUI often contain widgets of multiple types, levels, and states, and may change as the software is updated and changed. This requires that GUI widget recognition technology adapt to different GUI frameworks, platforms, and technologies, be flexible to deal with dynamic and asynchronous GUI elements, be compatible with different resolutions and scaling ratios, and deal with display issues like occlusion and overlap \cite{liu2020owl, yu2021layout}. 
Image-based recognition is easily affected by the image quality, similarity, deformation, \etc, which are prone to mismatching or missing matches. AI-based recognition requires a large amount of labeled data and computational resources. Finally, the integration of GUI widget recognition technology and testing tools is an issue. GUI widget recognition technology often needs to work in cooperation with other components to realize the end-to-end test automation process \cite{ran2022automated}. This requires testers to consider the matching degree and integration difficulty between the testing tool and the GUI widget recognition technology when selecting or using it.

GUI widget semantic analysis and understanding refers to understanding the meaning, function, and relationship of various GUI elements through specific technical means, so as to be able to generate, execute, and verify test cases more intelligently.

The GUI is composed of widgets of various types, levels, and states, and each widget has its own semantic information. 
The semantic understanding of the GUI interface involves inferring the relationship and function between the GUI widgets, and what operations and results users can get on the GUI interface. The information is very critical to verify whether the test cases are executed correctly and achieve the desired effect. However, due to the various possibilities and uncertainties on the GUI interface, and the differences in user operation habits and preferences, the semantic understanding of the GUI interface is a complex and fuzzy problem. Some existing approaches try to use rule-based, model-based, and learning-based methods for semantic understanding of GUI interfaces, but these methods still have some challenges, such as difficulty in covering all possible situations and scenarios, difficulty in handling exception and error situations, and difficulty in adapting to user feedback and changes in requirements \cite{yu2021prioritize}.

It is within these challenges that we find some opportunities for innovation and development. Through multimodal fusion, multiple types of information on the GUI interface can be processed at the same time and associations can be established between these information. It can also be combined with knowledge graph technology to contain various GUI elements and their relationships on the GUI interface. Knowledge graphs can also be used in combination with other AI techniques such as recommender systems, dialogue systems, \etc, to enable smarter, more personalized, and more adaptive testing. In addition, with the wide application of big data and cloud computing technologies, a large amount of GUI interface data can be collected and processed to provide richer and more diverse training samples. The elastic computing power of cloud computing can also be used to accelerate the training and inference process of the model to meet the real-time and large-scale testing requirements.

This part is trying to address a critical issue in vision-based GUI testing, spanning all the testing topics mentioned above. GUI element detection and semantic understanding are fundamental for automated test generation, test execution, and script maintenance. Existing element detection technologies still face challenges when dealing with complex and dynamic GUI, such as adapting to different platforms and systems, as well as handling asynchronous and dynamic GUI elements. We propose using multimodal fusion techniques to process various information on GUI interfaces, combined with knowledge graph technology, to enhance the accuracy and robustness of GUI element detection and semantic understanding. This provides technical support not only for GUI element detection but also serves as a foundation for different topics of GUI testing.

\subsection{Oracle Generation for GUI Testing}

In mobile app GUI testing, oracle plays a key role as a reference point for expected results. For humans, recognizing deviations from expected visual states is often intuitive. For example, if an image fails to load, human testers can immediately perceive it by observing blank areas or broken icons. However, machines need predefined oracles for accurate comparisons during automated testing, with these oracles serving as benchmarks during testing. In the process of buggy GUI element recognition, traditional CV techniques and deep learning models can be considered. However, it should be noted that these algorithms often focus on more prominent or outstanding content in the images. In contrast, the buggy GUI elements are always subtle, not easily observed, and often mixed with many similar elements \cite{yu2023mobile}. This characteristic makes them easy to be regarded as unimportant noise and filtered out during the recognition process. Therefore, directly using traditional algorithms in bug element detection makes it difficult to achieve ideal results \cite{yu2021prioritize}.

Future research can explore integrating semantic understanding into oracle generation. This involves enabling the system to understand the semantic meaning of visual elements in mobile apps. By understanding the context and purpose of each GUI element, machines can autonomously generate oracles that are not just simple visual matches. This can also extend to consideration of a broader context of the app, \ie, focusing on context-aware oracle generation. 
By understanding the context in which visual elements operate, oracles can be generated that capture subtle interactions within the app. Meanwhile, machine learning algorithms provide a path for the development of adaptive oracles, which can learn from historical results and user interactions and dynamically adjust oracle definitions according to the constant changes in apps. The adaptability is crucial for situations where GUI elements often change.

In addition, integrating user feedback into oracle generation brings the opportunity to improve the accuracy of expected results. It is supposed to be able to learn from user interactions and preferences and adjust oracles to meet user expectations. This bridges the gap between automated testing and actual user experience, making oracles more reflective of actual user expectations. As apps increasingly span multiple platforms, it is necessary to generate cross-platform compatible oracles. Research can explore methods for generating oracles compatible across different operating systems and devices to ensure the consistency and reliability of test results.

This part explores the issue of generating expected oracles during vision-based GUI testing, which is crucial for \gen, \rpt, \srp, and \eva. Existing visual recognition techniques and deep learning models often fall short in detecting subtle GUI bugs, especially in complex app scenarios. We propose introducing semantic understanding and context-aware technologies to enable GUI testing techniques to comprehend the semantics and interactions of GUI elements, thereby generating more accurate and adaptive test oracles.
In summary, by integrating advanced technologies such as machine learning, context awareness, user feedback, and cross-platform considerations, researchers are expected to develop adaptive and accurate oracles. This evolution is crucial for keeping up with the dynamism of modern apps and ensuring the effectiveness of automated vision-based GUI testing in various scenarios.

\subsection{Expert Domain Knowledge Integration for GUI Testing}

Expert domain knowledge refers to the knowledge about concepts, rules, methods, experiences, \etc, of a particular domain that is held by professionals in that domain. Expert domain knowledge integration in GUI testing refers to the use of different domain expert knowledge to guide and optimize the process and results of GUI testing by some technical means. The purpose of expert domain knowledge integration is to improve the effectiveness and efficiency of GUI testing, reduce the cognitive burden and decision-making difficulty of testers, and realize test automation and intelligence \cite{cacciotto2021metric, tuovenen2019mauto, zhu2021gui}.

There are some challenges in the integration of expert domain knowledge in GUI testing. The first is the acquisition and representation of expert domain knowledge. At present, some methods try to use ontologies, rules, cases, and so on to capture and represent expert domain knowledge. However, these methods still have some limitations, such as difficulty covering all possible expert domain knowledge, difficulty in dealing with incomplete and inconsistent expert domain knowledge, difficulty in updating and maintaining expert domain knowledge, \etc Secondly, the integration and application of expert domain knowledge is the core problem of expert domain knowledge integration, which involves how to establish connections and mappings between multiple different domains, how to abstract and refine between multiple different levels, and how to balance and optimize between multiple different goals. At present, some methods try to use model-based, learning-based, reasoning-based methods to integrate and apply expert domain knowledge. However, these methods still have some challenges, such as difficulty in dealing with conflict and coordination between multiple different domains, difficulty in dealing with consistency and traceability between multiple levels, and difficulty in dealing with balance and adaptability between multiple different goals.

Facing these challenges, we see opportunities, which motivate us to actively explore more innovative and intelligent ways to effectively integrate human expertise into GUI testing. For multi-domain conflict and coordination problems, we can consider adopting knowledge graphs to establish relationships by integrating knowledge from multiple domains, thus facilitating information flow across domains. In addition, creating a collaborative work platform is another way to solve these problems. This platform can make experts in different fields participate in knowledge integration and coordination together, and realize more organic cooperation. As for the problem of multi-level consistency and traceability, introducing a version control system is a feasible method, that helps to record and manage the evolution process of professional knowledge by tracking the changes in knowledge bases and models, ensuring the stability and reliability of the system. It is expected that the introduction of a series of innovative methods will overcome the problems faced by the current integration of expert domain knowledge, promote GUI testing in a more intelligent and efficient direction, and bring new development opportunities to the field of software quality assurance.

This part is trying to address the integration of expert domain knowledge into GUI testing, which is particularly important for improving the effectiveness of \gen, \frm, \rpt, and \ua. Current methods face challenges in knowledge acquisition, representation, and application, such as incomplete knowledge coverage, handling inconsistencies, and difficulties in updating and maintaining knowledge. We recommend employing knowledge graph technology and collaborative work platforms to facilitate the integration and coordination of multi-domain knowledge. These innovative methods can optimize the testing process, reduce cognitive burdens on testers, and enhance the automation and intelligence of testing.

\subsection{App Scenario Identification, Segmentation, Combination, and Understanding}

Mobile apps present a wide variety of scenarios covering a variety of user interactions and functionalities. Each scenario represents an organic set of user interactions, transitioning between multiple screens and performing a specific function. Strictly speaking, a test scenario is a specific function with complete and self-consistent business logic, consisting of a number of action pages or a number of code fragments.

However, the scenarios involved in modern apps often include different user actions, device states, network environments, \etc This makes it very complicated to identify, partition, and understand these diverse app scenarios, especially in large, complex apps. And, the combination of app scenarios can grow exponentially when multiple users, multiple devices, and multiple operations are involved, which makes testing coverage very difficult \cite{arlt2014reducing, wang2020combodroid}. In addition, app scenarios may change at different time points and under different conditions, increasing the uncertainty and dynamics of testing \cite{bae2012relative}. In the process of scenario recognition, the focus is on ensuring comprehensive coverage of potential operation paths. Future research can be devoted to developing intelligent algorithms that automatically identify key scenarios in mobile apps by combining CV and NLP technologies. For example, CV technology can effectively utilize GUI interface image information involved in traditional scenario exploration, making the testing scenario correspond to the GUI image and improving the accuracy of the description of the testing scenario. NLP technology can automatically generate or convert naturally described testing scenarios through training, or semantically understand and optimize classification of existing testing scenarios. Data mining techniques can also be used to deeply understand user navigation in apps, more accurately identify and associate app scenarios, and provide more targeted datasets for testing.

In terms of scenario segmentation, automatic segmentation algorithms can be explored by analyzing the page structure and code snippets of apps, intelligently dividing scenarios into manageable units. Such algorithms are expected to improve efficiency and accuracy during testing. Considering the dynamism of mobile apps, future research can also develop dynamic scenario segmentation methods. Techniques such as reinforcement learning can be used to adjust segmentation strategies in real-time to adapt to changes in apps during runtime. 

In addition, future research can focus on creating cross-scenario test suites to simulate processes executed by users in apps. This requires a deep understanding of interactions between different scenarios and the development of corresponding testing methods to ensure system robustness under complex user operations.

This part focuses on the identification, segmentation, and understanding of complex application scenarios. It addresses the issue of test coverage in large-scale applications in \gen, \rr, \frm, and \ua. We propose using computer vision and natural language processing techniques to automatically identify and generate testing scenarios. Additionally, we advocate for employing data mining techniques to gain deeper insights into user navigation behavior within applications and developing dynamic scene segmentation methods to adapt to changes during application runtime.

\subsection{Large Language Model Utilization in GUI Testing}

Large Language Models (LLMs) refer to high-capacity models with massive parameters and complex architectures used for deep learning tasks. They can understand and generate natural language text, as well as process multimodal information such as images, videos, audio, \etc In recent years, LLMs have achieved remarkable achievements in the field of NLP, such as GPT, LLaMA, \etc 
However, LLMs still have many challenges in being applied to GUI testing. The outputs of LLMs are often highly complex natural language expressions, and it can become very complicated for the automated approaches to understand and interpret these outputs. LLMs are usually pre-trained on a wide range of corpora, so their understanding may be relatively generalized and difficult to adapt to the testing requirements of a specific domain. In addition, designing test cases requires matching the output of the LLMs with the expected GUI behavior, which may involve efficient generation of test cases, automated execution, and verification of the results. Designing test cases with high coverage and capable of capturing diverse GUI behaviors is a non-trivial task.

Firstly, LLMs can utilize their powerful text generation capabilities to provide high-quality test cases, test scripts, test reports, \etc, for vision-based GUI testing. These text generation capabilities can improve the efficiency and effectiveness of testing and reduce the cost and errors of manual writing and maintenance \cite{feng2024prompting, liu2023fill}. 
The introduction of LLMs has the opportunity to gradually blur the boundaries of \gen and \rr because LLMs make it possible for automated exploration to also generate meaningful test scripts for testing migration, and the generated scripts are strong in logic. At the same time, the content explored and generated is also expected to be applied in scenario-based testing, such as scenario recognition, segmentation, combination, and understanding. The test scripts generated in these processes inevitably need maintenance and optimization when applied in practice. 

\begin{figure}[!htbp]
\centering
\vspace{-0.3cm}
\includegraphics[width=0.85\linewidth]{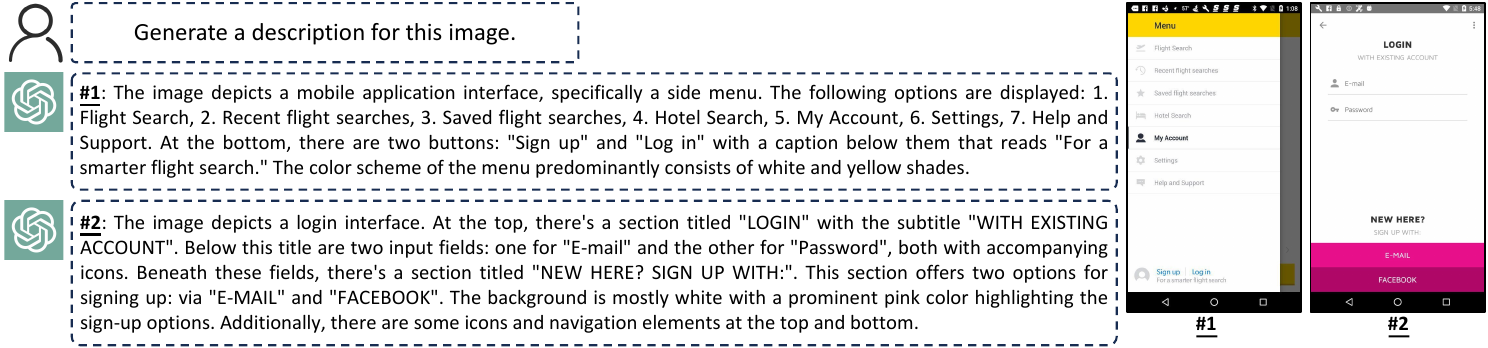}
\vspace{-0.3cm}
\caption{App GUI Screenshot Captioning with ChatGPT}
\vspace{-0.5cm}
\label{fig:llm}
\end{figure}

Secondly, LLMs can use their multimodal fusion capabilities to combine visual information and language information to achieve more in-depth and comprehensive testing of GUIs. These multimodal fusion capabilities can improve the coverage and accuracy of vision-based GUI testing and discover more potential defects and problems. For example, LLMs can judge whether there are problems such as layout confusion, color discordance, or unclear fonts in the GUI based on the screenshot of the GUI.
In \figref{fig:llm}, we show the caption generation capability of GPT-4(Vision) for app GUI screenshots, which is quite strong compared with non-LLM captioning models. GPT-4(Vision) can automatically identify the image is an app GUI screenshot and can precisely describe the app GUI contents. In addition, LLMs can also use their adaptive learning capabilities to conduct personalized and optimized testing based on different types and styles of GUIs. These adaptive learning capabilities can improve the flexibility and robustness of testing and adapt to different testing scenarios and requirements.


In summary, LLMs have broad application prospects and development space in GUI testing and are worth further exploration and research. LLMs not only provide more intelligent and efficient methods for vision-based GUI testing but also bring more innovation and value.

\section{Conclusion}

In this paper, we conduct a survey on the vision-based mobile app GUI testing techniques in the past years. Considering the dominant place of GUI testing in the whole lifecycle in mobile app development and maintenance, it is quite important to investigate the gradual development of GUI testing from code-based or layout-based to the current vision-based approaches. During the survey, we identify eight key topics for classifying GUI testing papers, covering the entire lifecycle of GUI testing, including generation, execution, maintenance, and analysis phases. Our survey covers different topics of mobile app GUI testing and we carefully discuss how CV technologies benefit the vision-based GUI testing techniques, which further enhance the mobile app GUI testing. Vision-based GUI testing techniques offer several advantages. First, such techniques can directly detect GUI elements from app screenshots to avoid issues related to missing elements in layout files, making the testing tools get rid of source code or layout file requirement. Second, such techniques adapt to different operating environments and devices, enhancing the generalizability of testing. Third, by detecting and locating GUI elements in app screenshots to execute user actions, the effectiveness of testing is improved. Despite the breakthroughs having been made by vision-based mobile app GUI testing, we further discuss the remaining challenges to be addressed in the future. Some limitations need to be solved or mitigated to help the vision-based techniques have better effectiveness and efficiency. We propose several points that should be studied on and can hopefully push forward the effectiveness and efficiency of vision-based mobile app GUI testing.

\begin{acks}
The authors would like to thank the anonymous reviewers for their insightful comments.
This work is supported partially by the National Natural Science Foundation of China (62141215, 62272220, 61802171), the Science, Technology and Innovation Commission of Shenzhen Municipality (CJGJZD20200617103001003), and the Fundamental Research Funds for the Central Universities (14380029).
\end{acks}

\bibliographystyle{ACM-Reference-Format}
\bibliography{main}

\newpage

\appendix

\section{Survey Methodology}
\label{sec:method}

\begin{figure}[!htbp]
\centering
\includegraphics[width=\linewidth]{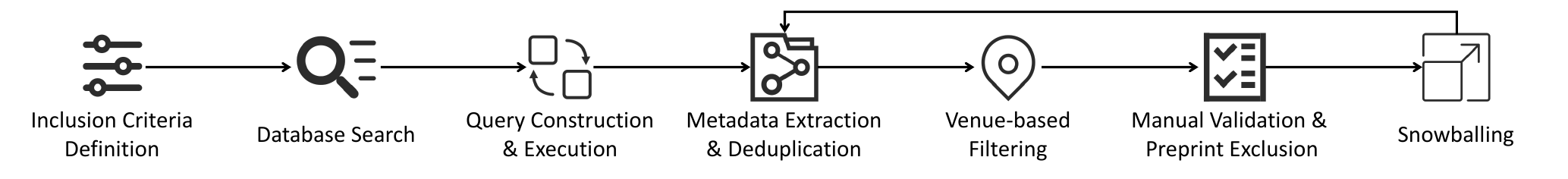}
\caption{Paper Collection Workflow}
\label{fig:collect}
\end{figure}

To ensure a comprehensive and objective survey of research in mobile app GUI testing, particularly focusing on vision-based approaches and the development trend, we adopt a systematic and transparent paper collection process. This process (as shown in \figref{fig:collect}) is designed to minimize human bias through the use of explicit inclusion criteria, automated scripts for metadata retrieval, and rigorous filtering and validation. Our aim is to construct a high-quality and representative corpus that accurately reflects the state of the field.

To begin, we define clear inclusion criteria to guide the selection process and ensure consistency. A paper is included if it satisfies at least one of the following conditions:

\begin{itemize}
	\item The paper introduces or discusses the general concepts and challenges of mobile app GUI testing.
	\item The paper proposes frameworks, tools, or methodologies specifically designed for testing mobile app GUI.
	\item The paper presents datasets or benchmarks explicitly tailored for mobile GUI testing.
	\item The paper offers technologies or techniques that directly assist the GUI testing process.
\end{itemize}

Conversely, we excluded studies that, while related to general software testing \cite{shahbazi2015black}\cite{yu2017test}, employed traditional methods with no GUI-specific considerations, for example, those that focused solely on logic-level testing or backend systems without addressing GUI elements.

To gather a wide and diverse set of relevant literature, we conduct keyword-based searches on two well-established databases: DBLP\footnote{\url{https://dblp.org/}} and Google Scholar\footnote{\url{https://scholar.google.com/}}. DBLP is chosen for its structured and comprehensive coverage of peer-reviewed computer science publications, while Google Scholar is included to broaden the search scope, particularly for early-access and influential studies that may not yet be indexed in DBLP. We do not further consider databases like Scopus because papers collected in such databases have been mainly covered by the selected databases. Our search is executed in October 2024 to ensure the corpus reflects the most up-to-date research available at the time.

We carefully construct three compound Boolean search queries to maximize recall across various terminology used in the field. These queries are: 

\begin{itemize}
	\item ((``GUI'' OR ``mobile app'' OR ``Android'' OR ``iOS'') AND ``testing'')
	\item ((``GUI testing'' OR ``mobile app testing'') AND (``screenshot'' OR ``image''))
	\item ((``GUI'' OR ``mobile'') AND (``testing'' OR ``app testing'') AND (``screenshot'' OR ``image''))
\end{itemize}

This approach is intended to capture both general studies on mobile GUI testing and more specific, vision-based techniques that leverage screenshots or visual data.

To reduce manual effort and improve reproducibility, we use a web crawler script to automatically retrieve metadata (\eg titles, author lists, venues, publication years, \etc) from the search results. This script operates within publicly available interfaces and adheres to the usage policies of both DBLP and Google Scholar. To handle the redundancy arising from query overlap, we apply an automated deduplication procedure that detects and removes duplicate entries based on exact matches of titles and authorship. This ensures the final dataset is clean and non-redundant before proceeding to the next stage.

After initial retrieval, we filter the results to include only papers published in six high-impact journals and fourteen prominent international conferences across software engineering, programming languages, and human-computer interaction. The venue list, detailed in \tabref{tab:venueyear}, is selected to ensure a high standard of peer review and topical relevance. We verify the publication status of all included papers and excluded preprints and any studies not formally accepted or published at the time of data collection.

To further expand our corpus and mitigate the limitations of keyword search, we apply snowballing \cite{wohlin2014guidelines} to each paper included in the initial result set. Specifically, we examine the related work sections and reference lists of these papers to identify additional studies that meet our predefined inclusion criteria. This iterative process allows us to capture papers that might use unconventional terminology or were otherwise missed in the original search.

\begin{figure}[!htbp]
\centering
\includegraphics[width=\linewidth]{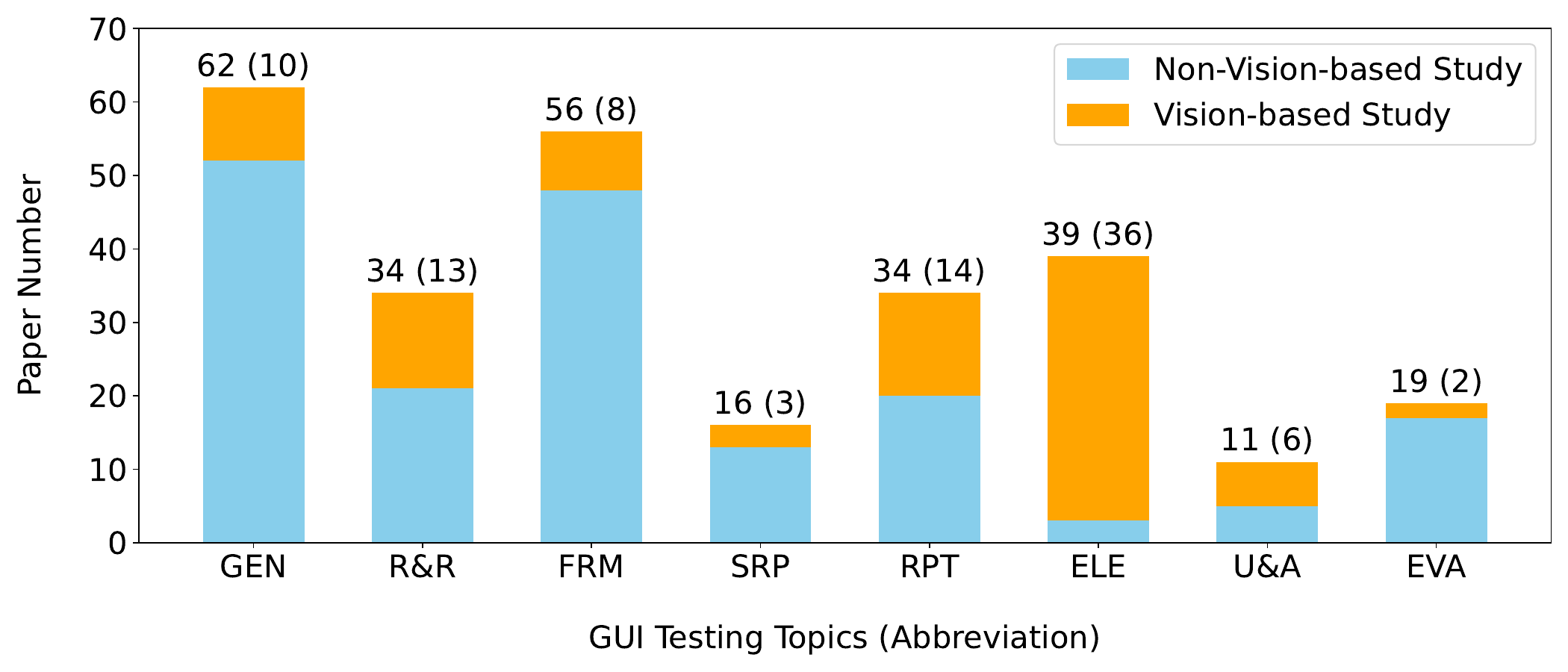}
\caption{Paper Number of Different Topics}
\label{fig:taxonomy}
\end{figure}

Through this multi-step process, we build up a final dataset of \papernum research papers. The topical distribution of these papers is illustrated in \figref{fig:taxonomy}. To ensure transparency and facilitate future replication, we have made the full list of included studies publicly available in our online repository: \textbf{\underline{\url{https://github.com/iGUITest/GUISurvey}}}.

Although the term vision-based is central to our focus, it was not included directly in the search strings due to its inconsistent usage across the literature and limited indexing in digital libraries. Instead, we constructed broad queries targeting mobile app GUI testing as a whole, ensuring that both traditional and emerging approaches were included. During the screening and classification phase, we manually identified vision-based techniques based on their reliance on visual inputs (e.g., screenshots, GUI images) as opposed to code-level or structural representations. This post hoc classification enables us to distinguish and analyze vision-based methods within the broader mobile GUI testing domain. While only a subset of the retrieved papers explicitly mention the term "mobile" in their titles or abstracts, we carefully reviewed the full texts to ensure relevance. Many papers target mobile-specific environments, frameworks (e.g., Android, iOS), or address challenges uniquely faced in mobile GUI testing, such as gesture interaction, device fragmentation, and platform constraints, even if the term "mobile" was not prominently stated. As a result, our dataset predominantly consists of mobile-focused work, and our classification reflects this alignment.

We have categorized the eight topics into three main categories: automated GUI testing, automation-assisted manual GUI testing, and fundamental techniques for GUI testing. This categorization is based on the evolution of GUI testing from manual testing to automated testing. The field of GUI testing has experienced a shift from complete reliance on manual operations to a gradual introduction of automation tools. While automated testing technologies can significantly enhance testing efficiency and coverage, human testing remains irreplaceable in certain complex scenarios. Therefore, current testing practices see a coexistence and mutual reinforcement of both human and automated testing. Additionally, both automated and human testing rely on various auxiliary testing tools that enhance the precision and effectiveness of testing. Our categorization not only reflects the development history of the GUI testing field but also illustrates the interdependence between different testing methods.

In terms of specific categorization, the automated GUI testing category includes \gen, \rr, and \frm. These topics focus on how to generate tests through automation, record and replay user actions in different environments, and provide a complete automated testing framework to support the entire testing process. Automated GUI testing can significantly reduce manual intervention, achieving more comprehensive and efficient test coverage. The \srp and the \rpt are categorized under the automation-assisted manual GUI testing category. These two topics mainly involve how to utilize automation tools to assist and optimize the manual testing process. Automation tools can help maintain and update testing scripts, ensuring they remain synchronized with the latest version of the apps, and can automatically extract key information through report analysis, reducing the workload of manual analysis. The fundamental techniques for GUI testing category includes \ele, \ua, and \eva. GUI element detection provides fundamental support for automated testing, enabling testing tools to accurately identify and manipulate GUI elements, \ie GUI widgets, while evaluation criteria are used to measure and compare the effectiveness and applicability of different testing techniques, providing an important reference for researchers and practitioners. GUI usability and accessibility testing is a fundamental topic that  guides the targets of both automated GUI testing and automation-assisted manual GUI testing.

Through this categorization, we aim to present the overall development trends in the GUI testing field and highlight the important roles of automation technologies and fundamental techniques for GUI testing in this process. Our categorization not only helps to understand the current state of research but also provides a clear direction for future studies.

\begin{figure}[!htbp]
\centering
\includegraphics[width=\linewidth]{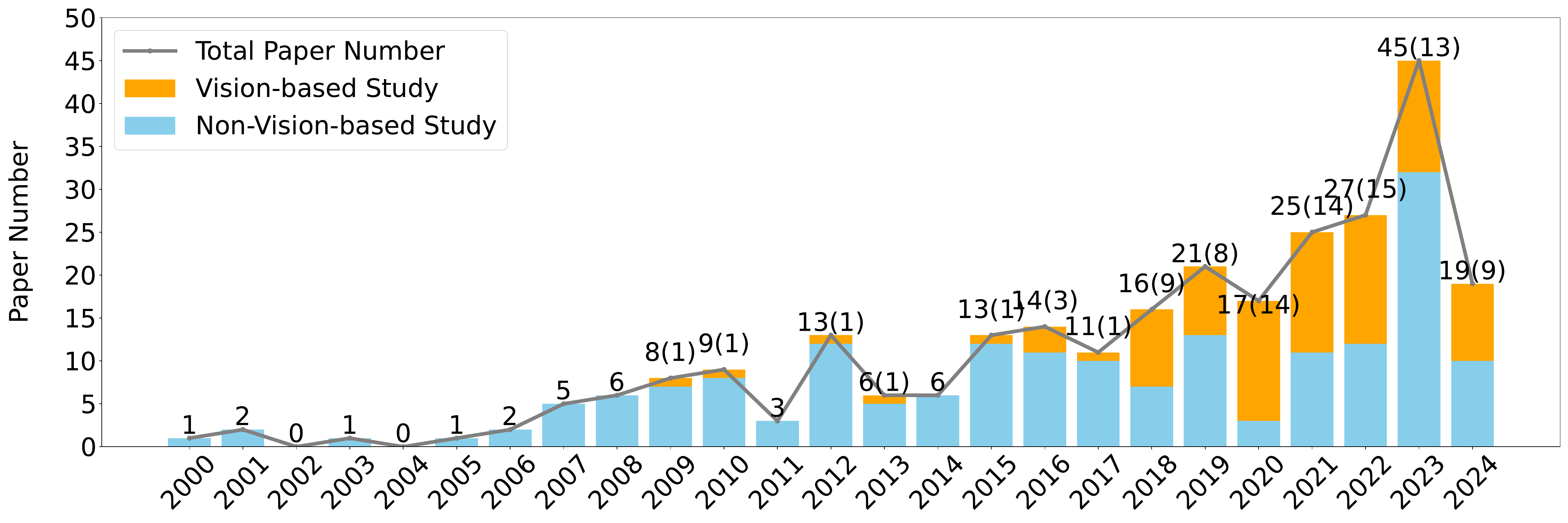}
\caption{Paper Number from 2000 to 2024}
\label{fig:year}
\end{figure}

The \gen topic has the most papers among different topics, which is 62, but the percentage of vision-based studies is not high. The \rr and \frm topics are also primary research topics in GUI testing, which have 34 and 56 papers, respectively. Within these two topics, there are some vision-based studies. For the \rpt topic, the percentage of vision-based studies is relatively higher because of a new testing paradigm, crowdsourced testing, which we discuss in \secref{sec:rpt}. There are many vision-based studies that focus on the \ele topic, which is supposed to be an important basis for the whole vision-based GUI testing. For the \srp, \ua and \eva topics, the numbers are 16, 11, and 19, respectively, with only a few vision-based studies. One possible reason is that these two topics are hard to be directly studied from a visual perspective.

\figref{fig:year} shows the number of studies from 2000 to 2024\footnote{Some venues have not published their 2024 proceedings since the paper collection of our survey. Our paper collection deadline is the end of September 2024.}. We can observe that the number of studies has increased dramatically since around 2012, indicating that GUI testing has attracted more and more attention since then. This trend corresponds to the development of mobile devices, mobile operating systems, mobile apps, \etc Since around 2016, the number of vision-based studies started to increase, which may have a relationship to the introduction of some advanced deep learning (DL) neural networks, \eg VGG-16 \cite{simonyan2014very}, ResNet \cite{he2016deep}, which significantly enhance the analysis and understanding to the app GUI.

\begin{figure}[!htbp]
\centering
\includegraphics[width=\linewidth]{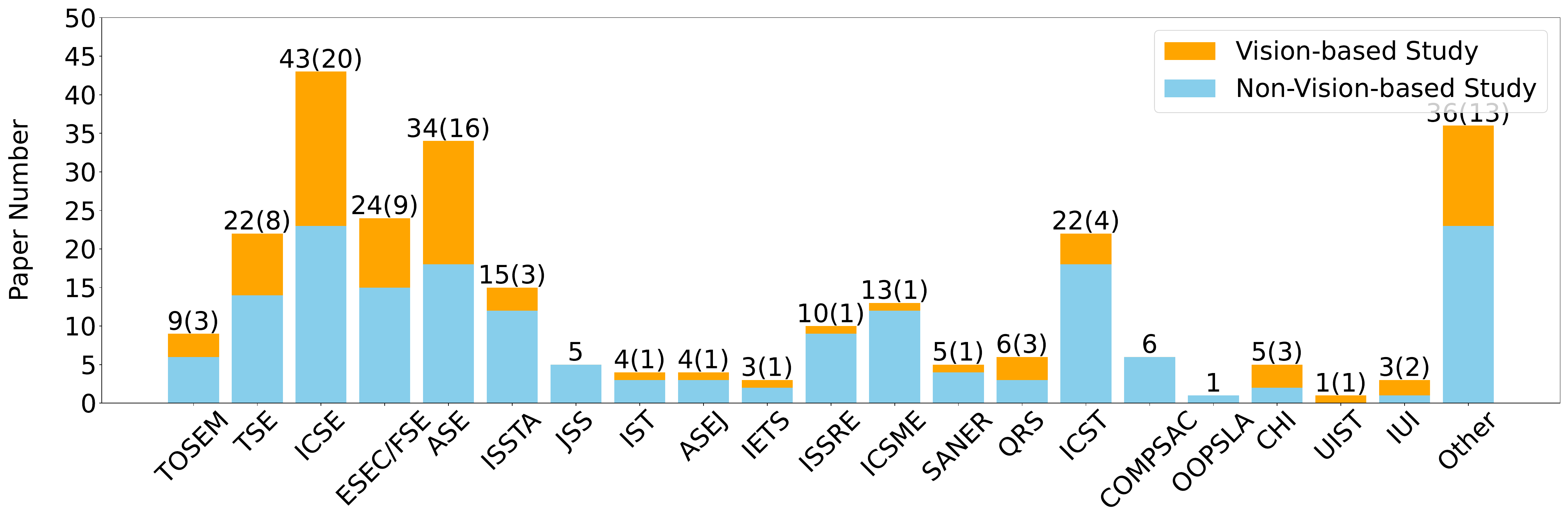}
\caption{Paper Number on Different Venues}
\label{fig:venue}
\end{figure}

\figref{fig:venue} shows the number of studies on different venues. Our paper collection involves different communities, including software engineering, programming language, and computer-human interaction. The programming language venue only has one paper published in OOPSLA 2015. For the human-computer interaction community, there are some publications in recent years, and they are mostly vision-based studies. Most publications are on the software engineering venues, and in the top-tier venues, like TSE, ICSE, and ASE, the percentage of vision-based studies is high.

\begin{table}[!htbp]
\centering
\caption{Venue-Year Matrix}
\scalebox{0.7}{
\begin{tabular}{c|ccccccccccccccccccccc|c}

\toprule
\rotatebox{90}{Venue} & 
\rotatebox{90}{TOSEM} & \rotatebox{90}{TSE} & 
\rotatebox{90}{ICSE} & \rotatebox{90}{ESEC/FSE} & 
\rotatebox{90}{ASE} & \rotatebox{90}{ISSTA} & 
\rotatebox{90}{JSS} & \rotatebox{90}{IST} & \rotatebox{90}{ASEJ} & \rotatebox{90}{IETS} & 
\rotatebox{90}{ISSRE} & \rotatebox{90}{ICSME} & \rotatebox{90}{SANER} & 
\rotatebox{90}{QRS} & \rotatebox{90}{ICST} & \rotatebox{90}{COMPSAC} & 
\rotatebox{90}{OOPSLA} & 
\rotatebox{90}{CHI} & \rotatebox{90}{UIST} & \rotatebox{90}{IUI} & 
\rotatebox{90}{\textit{Other}} & 
\rotatebox{90}{\textbf{Sum}} \\ \midrule

2000 & & & & & & \cellcolor{yellow}1 & & & & & & & & & & & & & & & & 1 \\
2001 & & \cellcolor{yellow}1 & & \cellcolor{yellow}1 & & & & & & & & & & & & & & & & & & 2 \\
2002 & & & & & & & & & & & & & & & & & & & & & & 0 \\
2003 & & & & & & & & & & & & \cellcolor{yellow}1 & & & & & & & & & & 1 \\
2004 & & & & & & & & & & & & & & & & & & & & & & 0 \\
2005 & & & & & & & & & & & & \cellcolor{yellow}1 & & & & & & & & & & 1 \\
2006 & & & & & & & & & & & \cellcolor{yellow}1 & \cellcolor{yellow}1 & & & & & & & & & & 2 \\
2007 & & & \cellcolor{yellow}1 & & \cellcolor{yellow}2 & & & & & & & \cellcolor{yellow}1 & & & & & & & & & \cellcolor{yellow}1 & 5 \\
2008 & \cellcolor{yellow}2 & & & & & & & & & \cellcolor{yellow}1 & & & & & & & & & & & \cellcolor{orange}3 & 6 \\
2009 & & & \cellcolor{yellow}2 & & & & & & & & & & & & \cellcolor{yellow}1 & \cellcolor{yellow}2 & & & \cellcolor{yellow}1 & & \cellcolor{yellow}2 & 8 \\
2010 & & \cellcolor{yellow}1 & \cellcolor{yellow}1 & & & & & & & & & & & & \cellcolor{orange}3 & & & \cellcolor{yellow}1 & & & \cellcolor{orange}3 & 9 \\
2011 & & \cellcolor{yellow}1 & & & \cellcolor{yellow}1 & & & & & & & & & & & & & & & & \cellcolor{yellow}1 & 3 \\
2012 & & & & & \cellcolor{orange}3 & & & & & \cellcolor{yellow}1 & \cellcolor{yellow}2 & \cellcolor{yellow}1 & & & \cellcolor{orange}3 & & & & & & \cellcolor{orange}3 & 13 \\
2013 & & & \cellcolor{yellow}1 & & & \cellcolor{yellow}1 & & & & & \cellcolor{yellow}1 & & & & \cellcolor{yellow}2 & & & & & & \cellcolor{yellow}1 & 6 \\
2014 & & & & & \cellcolor{yellow}1 & \cellcolor{yellow}2 & & & & & \cellcolor{yellow}1 & & & & & & & & & & \cellcolor{yellow}2 & 6 \\
2015 & & & & \cellcolor{yellow}1 & \cellcolor{orange}4 & & & & & & \cellcolor{yellow}2 & & & & \cellcolor{yellow}1 & \cellcolor{yellow}2 & \cellcolor{yellow}1 & & & & \cellcolor{yellow}2 & 13 \\
2016 & & \cellcolor{yellow}1 & \cellcolor{orange}4 & \cellcolor{yellow}1 & \cellcolor{orange}3 & \cellcolor{yellow}1 & & & & & & & & & \cellcolor{yellow}2 & & & & & & \cellcolor{yellow}2 & 14 \\
2017 & & & \cellcolor{yellow}1 & \cellcolor{yellow}2 & \cellcolor{orange}4 & & & & & & & \cellcolor{yellow}1 & & & \cellcolor{yellow}2 & & & & & & \cellcolor{yellow}1 & 11 \\
2018 & & \cellcolor{yellow}1 & \cellcolor{orange}3 & & \cellcolor{yellow}1 & & & & & & & & \cellcolor{yellow}1 & & \cellcolor{orange}3 & \cellcolor{yellow}1 & & & & & \cellcolor{red}6 & 16 \\
2019 & & & \cellcolor{orange}4 & & \cellcolor{orange}4 & \cellcolor{orange}3 & & \cellcolor{yellow}2 & & & \cellcolor{yellow}1 & \cellcolor{yellow}2 & \cellcolor{yellow}1 & \cellcolor{yellow}1 & & \cellcolor{yellow}1 & & & & & \cellcolor{yellow}2 & 21 \\
2020 & & \cellcolor{yellow}2 & \cellcolor{red}5 & \cellcolor{yellow}2 & \cellcolor{orange}3 & \cellcolor{yellow}1 & & & & \cellcolor{yellow}1 & & \cellcolor{yellow}1 & & \cellcolor{yellow}1 & & & & & & & \cellcolor{yellow}1 & 17 \\
2021 & \cellcolor{yellow}2 & & \cellcolor{red}5 & \cellcolor{orange}3 & \cellcolor{yellow}2 & \cellcolor{orange}4 & \cellcolor{yellow}1 & & & & & \cellcolor{yellow}1 & & \cellcolor{yellow}2 & \cellcolor{yellow}1 & & & \cellcolor{yellow}1 & & & \cellcolor{orange}3 & 25 \\
2022 & & \cellcolor{yellow}2 & \cellcolor{orange}3 & 7 & \cellcolor{orange}4 & \cellcolor{yellow}1 & & & & & & \cellcolor{yellow}1 & \cellcolor{yellow}1 & & \cellcolor{yellow}2 & & & \cellcolor{yellow}2 & & \cellcolor{yellow}1 & \cellcolor{orange}3 & 27 \\
2023 & \cellcolor{yellow}1 & \cellcolor{red}9 & \cellcolor{red}5 & \cellcolor{red}6 & \cellcolor{yellow}2 & \cellcolor{yellow}1 & \cellcolor{orange}3 & \cellcolor{yellow}2 & \cellcolor{orange}4 &  & \cellcolor{yellow}2 & \cellcolor{yellow}2 & \cellcolor{yellow}2 & \cellcolor{yellow}2 & \cellcolor{yellow}2 &  &  & \cellcolor{yellow}1 &  & \cellcolor{yellow}1 &  & 45 \\
2024 & \cellcolor{orange}4 & \cellcolor{orange}4 & \cellcolor{red}8 & \cellcolor{yellow}1 &  &  & \cellcolor{yellow}1 &  &  &  &  &  &  &  &  &  &  &  &  & \cellcolor{yellow}1 &  & 19 \\ \midrule

Sum & 9 & 22 & 43 & 24 & 34 & 15 & 5 & 4 & 4 & 3 & 10 & 13 & 5 & 6 & 22 & 6 & 1 & 5 & 1 & 3 & 36 & 271 \\ \bottomrule

\end{tabular}}
\label{tab:venueyear}
\end{table}

Further, we present a ``Venue-Year'' matrix in \tabref{tab:venueyear}, which illustrates the number of GUI testing studies on each venue of each year. The yellow, orange, and red colors represent 1-2 papers, 3-4 papers, and more than 4 papers, respectively. Intuitively, the vision-based GUI testing studies are concentrated in the left-bottom corner of the matrix, which represents the top-tier software engineering venues in recent 9 years. The venues that publish the most papers are ICSE and ASE.

\section{Paper Classification of Different Topics}

In this appendix section, we list the full paper approach clusterings of different topics.

In \tabref{tab:gen}, we focus on the \gen topic studies including random-based, model-based, system-based, learning-based approaches, together with the studies solving testing efficiency challenges in test generation.
\begin{table}[!htbp]
\centering
\caption{Taxonomy of GUI Test Generation}
\scalebox{0.9}{
\begin{tabular}{cc}

\toprule
Category & Approach \\ \midrule

Random-based &  
Kilincceker \etal \cite{kilincceker2019random}, 
Zeng \etal \cite{zeng2016automated} 
\\ \midrule

Model-based & \tabincell{c}{
Amalfitano \etal \cite{amalfitano2012using}, 
Arlt \etal  \cite{arlt2012lightweight}, 
Arlt \etal \cite{arlt2014reducing}, 
Brooks \etal \cite{brooks2007automated}, 
Ge \etal \cite{ge2023leveraging}, \\ 
Hu \etal \cite{hu2024enhancing}, 
Huang \etal \cite{huang2012apply}, 
Jensen \etal \cite{jensen2013automated}, 
Li \etal \cite{li2023human}, \\ 
Memon \etal \cite{memon2001hierarchical}, 
Mirzaei \etal \cite{mirzaei2015sig}, 
Ricos \etal \cite{ricos2023distributed}, 
Samuel \etal \cite{samuel2008automatic}, \\ 
Su \etal \cite{su2017guided}, 
Sun \etal \cite{sun2023property}, 
Tuglular \etal \cite{tuglular2009gui}, 
Wang \etal \cite{wang2022detecting}, \\ 
Wang \etal \cite{wang2020combodroid}, 
Wang \etal \cite{wang2023parallel}, 
Xie \etal \cite{xie2008using}, 
Yuan \etal \cite{yuan2007covering}, \\ 
Yuan \etal \cite{yuan2007using}, 
Yuan \etal \cite{yuan2009generating}, 
Zhang \etal \cite{zhang2017sketch} 
} \\ \midrule

System-based & \tabincell{c}{
Bae \etal \cite{bae2012relative}, 
Choudhary \etal \cite{choudhary2015automated}, 
Conroy \etal \cite{conroy2007automatic}, 
Gambi \etal \cite{gambi2023artisan}, \\ 
Li \etal \cite{li2009ontology}, 
Li \etal \cite{li2023crowdsourced}, 
Li \etal \cite{li2017droidbot}, 
Liu \etal \cite{liu2022guided}, 
Peng \etal \cite{peng2021cat}, \\ 
Win \etal \cite{win2023event}, 
Wu \etal \cite{wu2018sentinel}, 
Xie \etal \cite{xie2005rapid}, 
Yuan \etal \cite{yuan2008alternating} \\ 
} \\ \midrule

Learning-based & \tabincell{c}{
Beltramelli \etal \cite{beltramelli2018pix2code}, 
Carino \etal \cite{carino2015dynamically},
Guo \etal \cite{guo2023effectively}, 
Hu \etal \cite{hu2023appaction}, \\ 
Lan \etal \cite{lan2024deeply}, 
Li \etal \cite{li2019humanoid}, 
Liu \etal \cite{liu2023fill}, 
Liu \etal \cite{liu2024make}, 
Mariani \etal \cite{mariani2012autoblacktest}, \\ 
Pan \etal \cite{pan2020reinforcement}, 
Peng \etal \cite{peng2022mubot}, 
YazdaniBanafsheDaragh \etal \cite{yazdanibanafshedaragh2021deep}, \\ 
Yu \etal \cite{yu2024practical}, 
Yu \etal \cite{yu2024practicalnon},
Yu \etal \cite{yu2024effective}, 
Yu \etal \cite{yu2022universally}, 
Zhao \etal \cite{zhao2024dinodroid} 
} \\ \midrule

Testing Efficiency & \tabincell{c}{
Clerissi \etal \cite{clerissi2024guess}, 
Feng \etal \cite{feng2023efficiency}, 
Li \etal \cite{li2023crowdsourced}, \\ 
Pan \etal \cite{pan2023preference}, 
Wang \etal \cite{wang2021infrastructure}, 
Wang \etal \cite{wang2021vet}

} \\ \bottomrule

\end{tabular}}
\label{tab:gen}
\end{table}

 In \tabref{tab:rr}, we focus on the \rr topic studies including test script record and replay (migration) within the same app, cross different devices, cross different platforms, cross different apps, and specifically the event matching during test script record and replay (migration).
\begin{table}[!htbp]
\centering
\caption{Taxonomy of GUI Test Record \& Replay (Migration)}
\scalebox{0.9}{
\begin{tabular}{cc}

\toprule
Category & Approach \\ \midrule

Same-App & 
Amalfitano \etal \cite{amalfitano2019combining}, 
Feng \etal \cite{feng2022gifdroid}, 
Matos \etal \cite{matos2014record}, 
Zhao \etal \cite{zhao2022avgust} 
\\ \midrule

Cross-Device & \tabincell{c}{
Bernal \etal \cite{bernal2022translating}, 
Bernal \etal \cite{bernal2020translating}, 
Fazzini \etal \cite{fazzini2017barista}, 
Guo \etal \cite{guo2019sara},  \\ 
Halpern \etal \cite{halpern2015mosaic}, 
Havranek \etal \cite{havranek2021v2s}, 
Hu \etal \cite{hu2015versatile}, 
Li \etal \cite{li2022cross}, \\ 
Liu \etal \cite{liu2014capture}, 
Sahin \etal \cite{sahin2019randr}, 
Sahin \etal \cite{sahin2019towards}, 
Wu \etal \cite{wu2017appcheck}, 
} \\ \midrule

Cross-Platform & \tabincell{c}{
Alegroth \etal \cite{alegroth2013jautomate},  
Ji \etal \cite{vision2023ji}, 
Lin \etal \cite{lin2022gui}, 
Qin \etal \cite{qin2019testmig}, \\ 
Qin \etal \cite{qin2016mobiplay}, 
Steven \etal \cite{steven2000jrapture}, 
Talebipour \etal \cite{talebipour2021ui}, \\
Yu \etal \cite{yu2021layout},
Zhang \etal \cite{zhang2023resplay} 
} \\ \midrule

Cross-App & \tabincell{c}{
Behrang \etal \cite{behrang2018automated},  
El Ariss \etal \cite{el2010systematic},
Gomez \etal \cite{gomez2013reran},
\\ 
Lam \etal \cite{lam2017record}, 
Liang \etal \cite{liang2023rida}, 
Yu \etal \cite{yu2023llm}, 
Zhang \etal \cite{zhang2024learning} 
} \\ \midrule

\textit{Event Matching} & 
Lin \etal \cite{lin2018exploration}, 
Mariani \etal \cite{mariani2021semantic} 
\\ \bottomrule

\end{tabular}}
\label{tab:rr}
\end{table}

In \tabref{tab:frm}, we focus on the \frm topic studies including traditional-technology-based studies and innovation-approach-driven studies.
\begin{table}[!htbp]
\centering
\caption{Taxonomy of GUI Testing Framework}
\scalebox{0.9}{
\begin{tabular}{cc}

\toprule
Category & Approach \\ \midrule

\tabincell{c}{Traditional \\ Technology \\ -based} & \tabincell{c}{
Amalfitano \etal \cite{amalfitano2012toolset}, 
Bauersfeld \etal \cite{bauersfeld2013guidiff}, 
Bauersfeld \etal \cite{bauersfeld2012guitest}, 
Bello \etal \cite{bello2019opia}, \\
Bertolini \etal \cite{bertolini2010framework}, 
Cao \etal \cite{cao2019paraaim}, 
Cheng \etal \cite{cheng2016guicat}, 
Cheng \etal \cite{cheng2017systematic}, \\
Ermuth \etal \cite{ermuth2016monkey}, 
Feng \etal \cite{feng2007action}, 
Fischer \etal \cite{fischer2023insights}, 
Ganov \etal \cite{ganov2009event}, \\
Gao \etal \cite{gao2015pushing}, 
Gu \etal \cite{gu2019practical}, 
Guo \etal \cite{guo2020improving}, 
Li \etal \cite{li2014adautomation}, \\
Memon \etal \cite{memon2003dart}, 
Mirzaei \etal \cite{mirzaei2016reducing}, 
Moreira \etal \cite{moreira2013pattern}, 
Mu \etal \cite{mu2009design}, \\
Nguyen \etal \cite{nguyen2014guitar}, 
Ravelo \etal \cite{ravelo2019kraken}, 
Saddler \etal \cite{saddler2017eventflowslicer}, 
Song \etal \cite{song2017ehbdroid}, \\
Su \etal \cite{su2016fsmdroid}, 
Sun \etal \cite{sun2021setdroid}, 
Sun \etal \cite{sun2021understanding}, 
Verhaeghe \etal \cite{verhaeghe2019gui}, \\ 
Wen \etal \cite{wen2015pats}, 
Wetzlmaier \etal \cite{wetzlmaier2016framework}, 
Xie \etal \cite{xie2006model}, \\
Yao \etal \cite{yao2012distributed}, 
Zaraket \etal \cite{zaraket2012guicop}, 
Zhu \etal \cite{zhu2015context} 
} \\ \midrule

\tabincell{c}{Innovation \\ Approach \\ -Driven} & \tabincell{c}{
Cacciotto \etal \cite{cacciotto2021metric}, 
Chen \etal \cite{chen2022automatically}, 
Coppola \etal \cite{coppola2023effectiveness}, 
Eskonen \etal \cite{eskonen2020automating}, 
Gu \etal \cite{gu2017aimdroid}, \\ 
Hu \etal \cite{atom2023hu}, 
Kolthoff \etal \cite{kolthoff2023data}, 
Liang \etal \cite{liang2023ag3}, 
Liu \etal \cite{liu2023towards}, 
Liu \etal \cite{liu2022navidroid}, \\ 
Lv \etal \cite{lv2022fastbot2}, 
Moran \etal \cite{moran2016automatically}, 
Qian \etal \cite{qian2020roscript}, 
Ran \etal \cite{ran2022automated}, 
Rua \etal \cite{rua2023pyanadroid}, \\ 
Sun \etal \cite{sun2023characterizing}, 
Sun \etal \cite{sun2023lazycow}, 
Sun \etal \cite{sun2023taming}, 
Tuovenen \etal \cite{tuovenen2019mauto}, 
Wu \etal \cite{wu2023cydios}, \\ 
Yeh \etal \cite{yeh2009sikuli}, 
Yu \etal \cite{yu2016novel} 
} \\ \bottomrule

\end{tabular}}
\label{tab:frm}
\end{table}

In \tabref{tab:srp}, we focus on the \srp topic studies including test script generation, refactor, repair, and update.
\begin{table}[!htbp]
\centering
\caption{Taxonomy of GUI Test Script Maintenance}
\scalebox{0.9}{
\begin{tabular}{cc}

\toprule
Category & Approach \\ \midrule

Generation & 
Chang \etal \cite{chang2010gui}, 
Iyama \etal \cite{iyama2018automatically}, 
Yandrapally \etal \cite{yandrapally2014robust}
\\ \midrule

Refactor & 
Chen \etal \cite{chen2008gui}, 
Chen \etal \cite{chen2012bad}, 
Daniel \etal \cite{daniel2011automated} 
\\ \midrule

Repair & \tabincell{c}{
Cao \etal \cite{cao2024comprehensive}, 
Gao \etal \cite{gao2015sitar}, 
Grechanik \etal \cite{grechanik2009maintaining}, 
Huang \etal \cite{huang2010repairing}, 
Imtiaz \etal \cite{imtiaz2021automated}, \\
Memon \etal \cite{memon2008automatically}, 
Pan \etal \cite{pan2020gui}, 
Xu \etal \cite{xu2021guider}, 
Yoon \etal \cite{yoon2022repairing} 
} \\ \midrule

Update & 
Li \etal \cite{li2017atom}
\\ \bottomrule

\end{tabular}}
\label{tab:srp}
\end{table}

In \tabref{tab:rpt}, we focus on the \rpt topic studies including test script augmentation, captioning, clustering, duplication detection, prioritization, process assistance, quality detection, and reproduction.
\begin{table}[!htbp]
\centering
\caption{Taxonomy of GUI Test Report Analysis}
\scalebox{0.9}{
\begin{tabular}{cc}

\toprule
Category & Approach \\ \midrule

Augmentation & 
Moran \etal \cite{moran2016fusion} 
\\ \midrule

Captioning & 
Fang \etal \cite{fang2023test}, 
Feng \cite{feng2023read}, 
Liu \etal \cite{liu2018generating}, 
Yu \etal \cite{yu2019crowdsourced} 

\\ \midrule

Clustering & \tabincell{c}{
Cai \etal \cite{cai2021reports}, 
Cao \etal \cite{cao2020stifa}, 
Chen \etal \cite{chen2021effective}, \\ 
Du \etal \cite{du2022semcluster}, 
Jiang \etal \cite{jiang2018fuzzy},
Liu \etal \cite{liu2020clustering}, 
Yu \etal \cite{yu2024semi} 

} \\ \midrule

Duplicate Detection & \tabincell{c}{
Alipour \etal \cite{alipour2013contextual}, 
Banerjee \etal \cite{banerjee2012automated}, 
Cooper \etal \cite{cooper2021takes}, 
Hao \etal \cite{hao2019ctras}, \\ 
Jahan \etal \cite{jahan2023towards}, 
Jalbert \etal \cite{jalbert2008automated}, 
Nguyen \etal \cite{nguyen2012duplicate}, 
Sun \etal \cite{sun2011towards}, \\ 
Sun \etal \cite{sun2010discriminative}, 
Sureka \etal \cite{sureka2010detecting}, 
Wang \etal \cite{wang2019images} 
} \\ \midrule

Prioritization & 
Feng \etal \cite{feng2015test}, 
Feng \etal \cite{feng2016multi}, 
Yu \etal \cite{yu2021prioritize} 
\\ \midrule

Process Assistance & 
Fazzini \etal \cite{fazzini2022enhancing}, 
Song \etal \cite{song2023burt},  
Song \etal \cite{song2022toward} 
\\ \midrule

Quality Detection & 
Yu \etal \cite{yu2023mobile} 
\\ \midrule

Reproduction & 
Feng \etal \cite{feng2024prompting}, 
Huang \etal \cite{huang2024crashtranslator}, 
Zhao \etal \cite{zhao2022recdroid+}, 
Zhao \etal \cite{zhao2019recdroid} 
\\ \bottomrule

\end{tabular}}
\label{tab:rpt}
\end{table}

In \tabref{tab:ele}, we focus on the \ele topic studies including pure element detection tasks and domain application of element issues in app GUI.
\begin{table}[!htbp]
\centering
\caption{Taxonomy of GUI Element Detection}
\scalebox{0.9}{
\begin{tabular}{cc}

\toprule
Category & Approach \\ \midrule

Element Detection & \tabincell{c}{
Altinbas \etal \cite{altinbas2022gui}, 
Ardito \etal \cite{ardito2021feature}, 
Chen \etal \cite{chen2022towards}, 
Chen \etal \cite{chen2020object}, \\ 
Chiatti \etal \cite{chiatti2018text}, 
De \etal \cite{de2024investigating}, 
Fang \etal \cite{fang2016automatic}, 
Feiz \etal \cite{feiz2022understanding}, \\ 
Jaganeshwari \etal \cite{jaganeshwari2021novel}, 
Ji \etal \cite{ji2018uichecker}, 
Kirinuki \etal \cite{kirinuki2022web}, \\ 
Kolthoff \etal \cite{kolthoff2023data}, 
Liu \etal \cite{liu2020discovering}, 
Liu \etal \cite{liu2022nighthawk}, 
Mahmud \etal \cite{mahmud2024using}, \\ 
Mansur \etal \cite{mansur2023aidui}, 
Moran \etal \cite{moran2018automated}, 
Moran \etal \cite{moran2018detecting}, 
Nie \etal \cite{nie2024sok}, \\ 
Sun \etal \cite{sun2020ui}, 
White \etal \cite{white2019improving}, 
Xie \etal \cite{xie2020uied}, \\ 
Zhang \etal \cite{zhang2021construction}, 
Zhang \etal \cite{zhang2021screen} 
} \\ \midrule

Domain Application & \tabincell{c}{
Abdelhamid \etal \cite{abdelhamid2020deep}, 
Bajammal \etal \cite{bajammal2018web}, 
Chen \etal \cite{chen2018ui}, \\ 
Hu \etal \cite{hu2023automated}, 
Leiva \etal \cite{leiva2022describing}, 
Moran \etal \cite{moran2018machine}, 
Nass \etal \cite{nass2023robust}, \\ 
Nguyen \etal \cite{nguyen2015reverse}, 
Qian \etal \cite{qian2022accelerating}, 
Xiao \etal \cite{xiao2019iconintent}, 
Xie \etal \cite{xie2022psychologically}, \\ 
Ye \etal \cite{ye2021empirical}, 
Zhang \etal \cite{zhang2022unirltest}, 
Zhao \etal \cite{zhao2021guigan}, 
Zhu \etal \cite{zhu2021gui} 
} \\ \bottomrule

\end{tabular}}
\label{tab:ele}
\end{table}

In \tabref{tab:ua}, we have a special focus on the \ua topic studies.
\begin{table}[!htbp]
\centering
\caption{Taxonomy of GUI Usability / Accessibility Testing}
\scalebox{0.9}{
\begin{tabular}{cc}

\toprule
Category & Approach \\ \midrule

Usability / Accessibility Issue & \tabincell{c}{
Alshayban \etal \cite{alshayban2022accessitext}, 
Chen \etal \cite{chen2020unblind}, 
Chen \etal \cite{chen2021accessible}, 
Hu \etal \cite{hu2023exploring}, \\ 
Liu \etal \cite{liu2020owl}, 
Salehnamadi \etal \cite{salehnamadi2022groundhog}, 
Su \etal \cite{su2022metamorphosis}, 
Tazi \etal \cite{tazi2023accessibility}, \\ 
Zhang \etal \cite{zhang2023accessfixer}, 
Zhang \etal \cite{zhang2023automated}, 
Zhao \etal \cite{zhao2020seenomaly} 
} \\ \bottomrule

\end{tabular}}
\label{tab:ua}
\end{table}

In \tabref{tab:eva}, we focus on the \eva topic studies including comparison evaluation of GUI testing techniques, evaluation criteria, and other evaluation topics.
\begin{table}[!htbp]
\centering
\caption{Taxonomy of GUI Testing Evaluation Criteria}
\scalebox{0.9}{
\begin{tabular}{cc}

\toprule
Category & Approach \\ \midrule

Comparison Evaluation & \tabincell{c}{
Alegroth \etal \cite{alegroth2015conceptualization}, 
Amalfitano \etal \cite{amalfitano2015conceptual}, 
Bertolini \etal \cite{bertolini2010gui}, \\
Bertolini \etal \cite{bertolini2009empirical}, 
Bons \etal \cite{bons2023scripted}, 
Borjesson \etal \cite{borjesson2012automated}, 
Coppola \etal \cite{coppola2018mobile}, \\
Dobslaw \etal \cite{dobslaw2019estimating}, 
Su \etal \cite{su2021benchmarking} 
} \\ \midrule

Evaluation Criteria & \tabincell{c}{
Baek \etal \cite{baek2016automated}, 
Elyasaf \etal \cite{elyasaf2023generalized}, 
He \etal \cite{he2015gui}, \\
Ma \etal \cite{ma2023automata}, 
Memon \etal \cite{memon2001coverage}, 
Yuan \etal \cite{yuan2010gui} 
} \\ \midrule

\textit{Other} & \tabincell{c}{
Alegroth \etal \cite{alegroth2018continuous}, 
Jaaskelainen \etal \cite{jaaskelainen2009automatic}, \\
Rauf \etal \cite{rauf2010automated}, 
Xie \etal \cite{xie2006studying}
} \\ \bottomrule

\end{tabular}}
\label{tab:eva}
\end{table}

\section{LLM Application in Different GUI Testing Topics}

LLMs can significantly enhance mobile app GUI testing by visually understanding and interpreting the GUI information and semantics. Leveraging advanced computer vision techniques, LLMs can analyze the layout, components, and interactions within a mobile app GUI, identifying elements such as buttons, icons, text fields, and navigation bars. This allows LLMs to simulate user actions, verify the correctness of visual components, and detect inconsistencies or defects, such as misalignments or broken links. By automating these tasks, LLMs assist in streamlining testing processes, improving accuracy, and reducing the need for manual intervention to ensure an app's visual integrity and functionality.

As with the eight topics mentioned above, LLM can have an effect on all the topics. LLMs are expected to enhance the \gen through their capabilities to understand GUI elements and natural language descriptions. LLMs can autonomously explore different pathways in apps, simulating various user behaviors and discovering untested scenarios. LLMs can generate exploratory tests that go beyond predefined workflows, identifying hidden bugs. LLMs can dynamically adapt their exploration based on real-time feedback, such as detecting unexpected behavior or GUI changes, ensuring that all aspects of the GUI are tested. This intelligent, adaptive approach enhances test coverage, especially for complex applications, while reducing manual effort in creating exploratory tests. 

Besides, LLMs can enhance the process of \rr. During the recording phase, LLMs can intelligently interpret user interactions with the mobile app GUI, automatically identifying and categorizing GUI elements like buttons, input fields, and dropdowns. They can generate human-readable scripts that precisely capture these interactions, making it easier for testers to understand and modify them. For the replay phase, LLMs can adapt to subtle changes in the app GUI layout or workflow, such as repositioned buttons or updated labels, ensuring that the test scripts remain robust even after UI updates. 

Additionally, LLMs can enhance \frm by improving both test execution and management. In the execution phase, LLMs can automatically adapt to variations in app interfaces, identifying changes such as updated buttons or altered layouts and adjusting test scripts dynamically to ensure smooth execution without manual intervention. In terms of test management, LLMs can streamline processes by generating reports, categorizing test results, and providing actionable insights on test coverage, defect identification, and areas needing further testing. 

Regarding the \srp, LLMs can simplify and enhance the process by intelligently adapting to changes in the app GUI. As mobile apps evolve with frequent updates and GUI modifications, traditional test scripts often break or require manual adjustments. LLMs can automatically detect these changes and update the test scripts accordingly, ensuring they remain functional. LLMs can analyze patterns in previous test failures, suggesting fixes or optimizations for future test scripts. This reduces the burden on developers and testers by minimizing manual intervention, lowering maintenance costs, and increasing the reliability and longevity of test scripts even in rapidly evolving app environments. 

Furthermore, LLMs can enhance the \rpt analysis by automatically interpreting complex test results and providing actionable insights. When analyzing large volumes of data from automated tests, LLMs can quickly identify patterns, detect recurring issues, and highlight critical failures, reducing the time testers spend sifting through detailed logs. They can provide natural language summaries of test outcomes, prioritizing issues that impact functionality or user experience. LLMs can correlate test failures with recent changes in the app, offering suggestions for troubleshooting and fixes. 

In terms of \ele, LLMs can improve by leveraging advanced natural language processing and computer vision capabilities to accurately identify and interpret a wide range of UI components. Traditional methods often rely on predefined identifiers or static attributes, but LLMs can understand the context and intent behind elements like buttons, text fields, and icons, even when they vary across different screens or versions of the app. By recognizing both visual and semantic cues, LLMs can detect GUI elements more flexibly, handling changes in layout, style, or functionality without the need for manual updates. This adaptability enhances the accuracy of automated tests and user interaction simulations, reducing false positives or missed GUI elements and ensuring a more robust and resilient testing process. 

In terms of \ua, LLMs can enhance by providing intelligent analysis of the app GUI based on established usability and accessibility guidelines. For usability testing, LLMs can evaluate the app's design for clarity, consistency, and ease of navigation, identifying potential pain points such as confusing layouts or difficult-to-find features. In terms of accessibility, LLMs can automatically assess the app against standards, detecting issues such as insufficient color contrast, missing alt text, or non-compliant font sizes. Moreover, LLMs can simulate interactions from the perspective of users with disabilities, such as screen reader users or those with motor impairments, ensuring that the app is inclusive and user-friendly for all. By automating these evaluations, LLMs help developers quickly identify and address usability and accessibility issues, leading to a better user experience for a diverse audience. 

Last but not least, LLMs can enhance \eva by automating the assessment of test results and providing deeper insights into the app performance and GUI quality. LLMs can evaluate the effectiveness of test cases by analyzing test coverage, identifying areas of the app that may not have been thoroughly tested, and recommending additional tests to improve coverage. They can also interpret complex test data, identifying patterns of failures or inconsistencies in the user interface, such as layout misalignments or broken elements across different device configurations. Furthermore, LLMs can offer real-time feedback on the app responsiveness, performance, and user experience, enabling testers to pinpoint issues faster and more accurately.

\section{Discussion of Potential Threats}

\subsection{Approach Basis Classification}

Using vision-based techniques in mobile app GUI testing is the main focus of this research. During the classification of collected studies, we acknowledge that subjectivity inevitably exists in the process of distinguishing between visual and non-visual studies. However, we have tried our best to mitigate this threat. First, we define that vision-based GUI studies primarily utilize visual information, such as image processing, computer vision, or screenshot analysis. Non-vision-based studies, on the other hand, encompass those that focus on traditional software testing methods without heavy reliance on visual data, for instance, code analysis or event-flow testing \cite{shahbazi2015black, yu2017test}. Second, to further reduce subjectivity, the initial classification was conducted by three authors independently and reviewed by two other authors. Discrepancies were resolved through discussion. Finally, a classification consistency test was conducted on the papers to evaluate the agreement index among different authors, ensuring the reliability of the classification.

\subsection{Paper Selection Process}

To ensure comprehensive topic coverage and reduce selection bias, we employed a two-phase strategy: (1) systematic database search using defined inclusion criteria (as outlined in \secref{sec:method}), and (2) snowballing by reviewing the related work sections of the initially retrieved studies. This iterative approach allowed us to identify additional relevant papers beyond those captured by the initial keyword queries.

Our primary focus is on mobile app GUI testing. Accordingly, we concentrated on papers published in six journals and fourteen top-tier conferences that are recognized as major venues in software engineering, programming languages, and human-computer interaction. These venues represent the most active and visible forums for research in our target area.

Regarding the role of AI/ML/DL-based research, we made a conscious decision to exclude studies that propose general-purpose machine learning methods without directly addressing GUI testing challenges. While such methods may provide foundational techniques, our goal is to survey applied work that tackles the unique problems of mobile app GUI testing. When AI/ML/DL methods are used specifically to enhance GUI testing, for example, in screenshot classification or visual bug detection, they are included in our corpus.

By combining systematic search, iterative expansion, and explicit documentation of inclusion boundaries, we aim to ensure a diverse and representative survey corpus while maintaining a clear and focused scope.

\end{document}